\documentclass[%
preprint,
 amsmath,amssymb,
 aps,
]{revtex4-2}
\usepackage{siunitx}
\usepackage{graphicx}
\usepackage{dcolumn}
\usepackage{bm}
\usepackage{xcolor}

\newcommand{\myparallel}{{\mkern3mu\vphantom{\perp}\vrule depth 0pt\mkern2mu\vrule depth 0pt\mkern3mu}}

\begin{document}

\preprint{APS/123-QED}
\title{
Domain-engineered ferroelectric BiFeO$_3$ thin films for efficient bias-free THz emission}

\author{Z. Abdul Hadi$^{1}$, N. Rezi$^{2}$, M. Monti$^{1}$, G. Vaudel$^{1}$,  A. Abdelsamie$^{2}$, C. Carrétéro$^{2}$, D. Sando$^{3,4}$, M. Viret$^{5}$, V. Garcia$^{2}$, S. Fusil$^{2}$, M. C. Weber$^{1}$, P. Ruello$^{1}$, V. Juv\'e $^{1}$\footnote{Electronic address: vincent.juve@univ-lemans.fr}}

\affiliation{
$^{1}$ Le Mans Université, CNRS, IMMM, Institut des Molécules et des Matériaux du Mans, UMR 6283, F-72000 Le Mans, France\\
$^{2}$ Laboratoire Albert Fert, CNRS, Thales, Universit\'e Paris-Saclay, 91767 Palaiseau, France\\
$^{3}$ The MacDiarmid Institute for Advanced Materials and Nanotechnology, School of Physical \& Chemical Sciences, University of Canterbury, New Zealand\\
$^{4}$ School of Materials Science and Engineering, UNSW Sydney, Australia\\
$^{5}$ SPEC UMR CEA/CNRS, Université Paris-Saclay, L’Orme les Merisiers, 91191 Cedex, Gif-sur-Yvette, France\\
}

\date{\today}

\begin{abstract} 
\textbf{Ultrashort light pulses can drive transient photocurrents in solids, the basis of modern terahertz (THz) emitters. Ferroelectrics have recently emerged in this context as sources that require neither the bias voltage of photoconductive antennas nor the magnetic field of spintronic emitters, yet the ultrafast photocurrent generating the emission has remained debated. Here, we resolve this question in epitaxial BiFeO$_3$ thin films with engineered ferroelectric domain configurations, from single-domain to periodic stripe patterns. The stripe-domain film emits a THz field four times stronger than the single-domain configurations, reaching about half the output of a metallic spintronic reference emitter while remaining unsaturated at our highest excitation fluences. Resolving the emitted THz field in amplitude and phase as a function of the pump light polarization and the crystal orientation, we separate the coexisting photocurrents by their symmetry. This identifies the ultrafast screening of the ferroelectric polarization by photocarriers as the dominant emission mechanism in the in-plane-polarized films, enhanced in multidomain films by the strong built-in fields at the domain walls, while an out-of-plane polarization favors the bulk photovoltaic current. Our results establish the ferroelectric domain configuration as a design parameter setting both the strength and the microscopic origin of light-driven THz emission.}
\end{abstract}

\maketitle
\section{Main}

Ultrashort laser pulses directly excite charge and spin carriers in condensed matter, and the resulting non-equilibrium distributions generate transient currents that emit electromagnetic radiation in the terahertz (THz) range. Beyond its use as a probe of ultrafast electronic and spintronic processes~\cite{Leitenstorfer2023}, this emission provides strong THz sources~\cite{Meineke2022,Rouzegar2023}. Photoconductive switches, pioneered by Auston and colleagues, combine femtosecond excitation with an external electric field to produce ultrafast current transients in semiconductors~\cite{Zhang1990,ShanJieandHeinz2004}. More recently, spintronic emitters have exploited ultrafast spin-to-charge conversion in ferromagnetic/non-magnetic heterostructures, with an external magnetic field setting the magnetization~\cite{Kampfrath2013,seifert2016,Seifert2022,Micica2025}. Compared with optical rectification in bulk nonlinear crystals (e.g., ZnTe, LiNbO$_3$)~\cite{Yeh2007,Hirori2011,Fulop2020}, such thin-film emitters require no phase matching, easing the design of compact and integrated THz devices, and their bias fields offer straightforward modulation of the emission. However, the very requirement for these external fields introduces significant hurdles for integration and miniaturization in complex device environments.

A central challenge for future THz technologies is thus the realization of thin-film emitters operating efficiently without external electric or magnetic fields, from photoinduced currents with tunable amplitude, polarization, and directivity. These requirements direct attention toward multiferroic materials, and in particular toward bismuth ferrite (BiFeO$_3$), a single-phase multiferroic with robust ferroelectric and antiferromagnetic order at room temperature, widely investigated for nanoelectronic, photovoltaic, and magnetoelectric applications~\cite{Catalan2009,Heron2014,Sando2014,Paillard2016}. Its large spontaneous polarization ($\approx 100~\mu\mathrm{C/cm}^2$~\cite{Catalan2009}) supports several possible sources of bias-free THz emission: light-induced shift currents (the bulk photovoltaic effect), ultrafast acceleration of photocarriers by the internal depolarizing field, and rapid changes in the magnetic order. Moreover, specific ferroelectric domain patterns, from single-domain to stripe configurations, can be stabilized by epitaxial strain through substrate choice and growth conditions~\cite{Sando2016}, enabling the tuning of the electrical, magnetic, and optical responses~\cite{Sando2016,Paillard2016,Chauleau2020,Haykal2020,Sando2022} and making BiFeO$_3$ thin films attractive candidates for bias-free THz emission.


Despite this potential, experimental investigations of THz emission from BiFeO$_3$ remain scarce, and the underlying emission mechanisms are still debated~\cite{Takahashi2006,Takahashi2007,Talbayev2008,Rana2009,Guzelturk2020}. Two gaps stand out: a quantitative comparison of the emission efficiency with benchmark emitters, such as spintronic FeCoB/Pt bilayers, and the influence of the ferroelectric domain orientation and patterning on the emission, a question made intricate by the tensorial nature of the light-induced effects and the anisotropies associated with ferroelectricity. To address both issues, and making use of recent advances in strain engineering~\cite{Chauleau2020,Dufour2023,Gu2023}, we systematically compare the THz emission from BiFeO$_3$ films with different ferroelectric configurations: single-domain states with polarization oblique or normal to the surface, and multidomain patterns exhibiting a macroscopic net polarization.

We demonstrate that domain-engineered BiFeO$_3$ thin films achieve robust, bias-free, and efficient THz emission at room temperature, with the multidomain films competing with spintronic emitters such as FeCoB/Pt~\cite{seifert2016} at high light excitation density. Decomposing the complex THz field into its angular harmonics, we identify the ultrafast screening of the ferroelectric polarization by photocarriers as the dominant emission mechanism in the in-plane-polarized films, while the out-of-plane-polarized film is governed by the bulk photovoltaic effect, and show that the multidomain configuration enhances the screening current through the uncompensated internal fields of its domain walls. THz emission is thus set by the ferroelectric domain organization, opening avenues for tunable, compact, bias-free THz devices through domain engineering.

\begin{figure}[]
\includegraphics[width=\columnwidth]{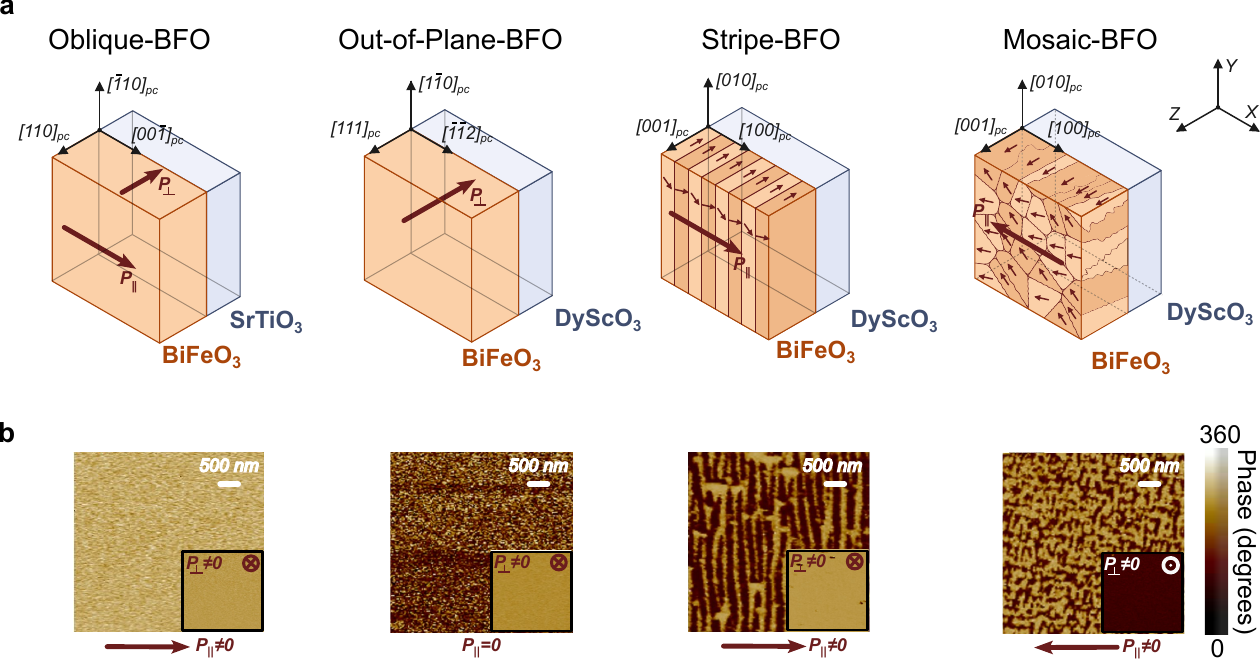}
\caption{\label{fig:fig1} \textbf{Orientation and domain configurations of the BiFeO$_3$ thin films.}
(a) Schematic of the BiFeO$_3$ thin films showing their crystallographic axes and
ferroelectric polarization orientations: the large arrows denote the polarization,
with $\mathbf{P}_\myparallel$ and $\mathbf{P}_{\perp}$ the in-plane and out-of-plane components with
respect to the sample surface. The buffer layers are omitted for clarity. For
Stripe-BFO and Mosaic-BFO, only the macroscopic in-plane ferroelectric polarization
is shown, arising from the combined contributions of the individual domain
polarizations (small arrows). $(X,Y,Z)$ is the crystal coordinate system. (b) Lateral PFM images of the different samples. Vertical PFM images of the different samples are shown as insets.}
\end{figure}

\begin{figure}[]
\includegraphics[width=\columnwidth]{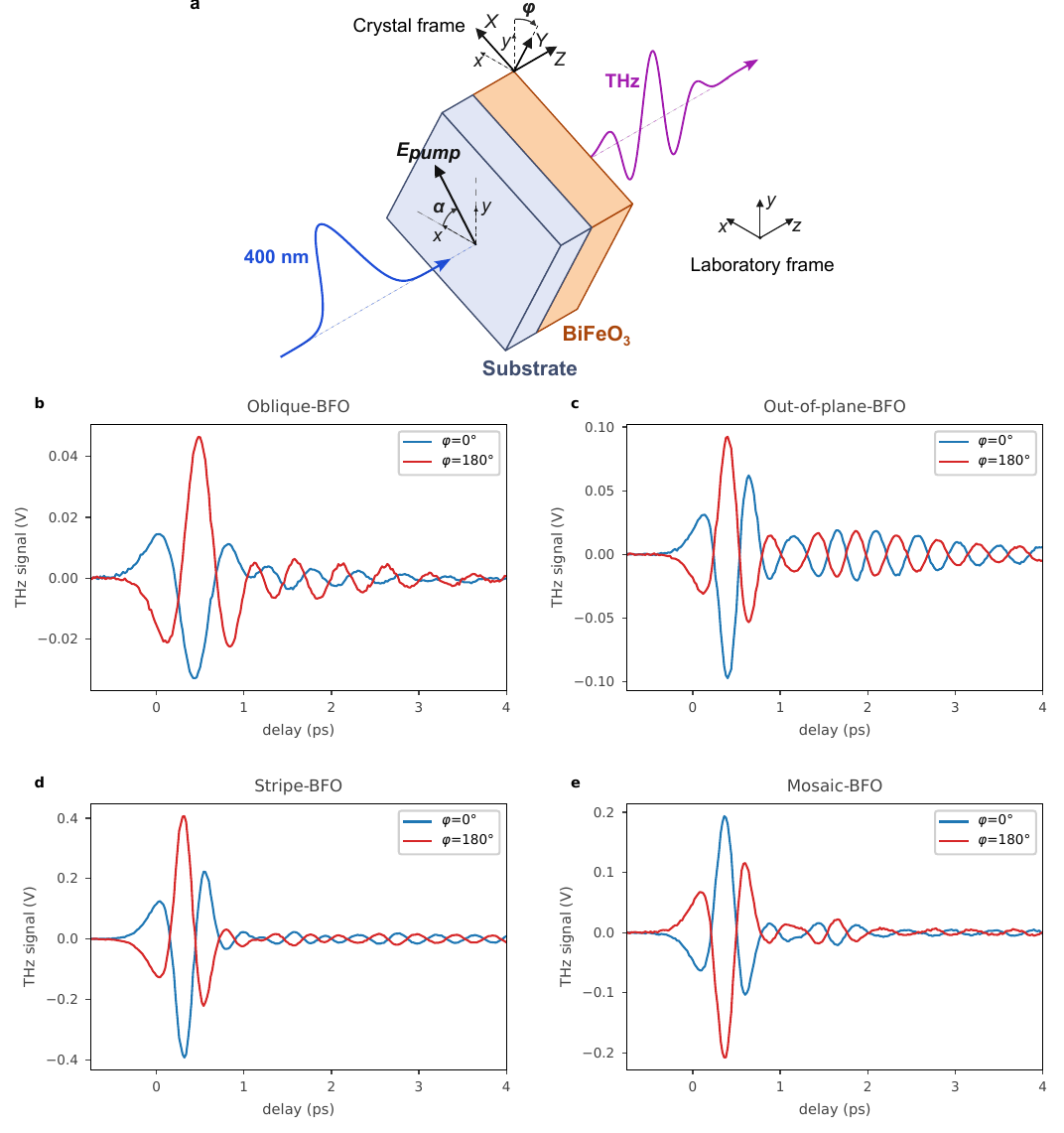}
\caption{\label{fig:fig2} \textbf{THz emission from BiFeO$_3$ thin films excited by an ultrafast optical pulse.}
(a) Experimental configuration for generating the THz pulse. The pump electric field
($\vec{E}_\mathrm{pump}$) polarization angle ($\alpha$) and the sample azimuthal
orientation angle ($\varphi$) are defined with respect to the laboratory $(x,y)$ plane.
$(X,Y,Z)$ is the crystal coordinate system as defined in Fig.~\ref{fig:fig1} for each thin films.
(b--e) THz waveforms generated for each thin film. The pump light pulse is polarized along the $y$-axis ($\alpha=90^\circ$) for either an azimuthal $\varphi$ angle set to $0^\circ$ (blue) or $180^\circ$ (red), while the THz electric field polarization component projected along the laboratory $x$-axis is detected. Note the differing $y$-axis scales.}
\end{figure}

\section{Results and discussion}

For a direct evaluation of THz emission from thin films and of the underlying mechanisms, we select four types of epitaxial BiFeO$_3$ films with distinct ferroelectric domain configurations (Fig.~\ref{fig:fig1}; see Methods and Supplementary Section I for growth, structure, and experimental details). Oblique-BFO (180\,nm thick) is single-domain with its polarization oblique to the surface, its in-plane component $\mathbf{P}_\myparallel$ lying along $[00\bar{1}]_\textrm{pc}$. Out-of-plane-BFO (125\,nm thick) is single-domain with the polarization essentially perpendicular to the surface, and exhibits an anisotropic in-plane strain along $[\bar{1}\bar{1}2]_\textrm{pc}$~\cite{Dufour2023}. Stripe-BFO (100\,nm thick) hosts a periodic pattern with 71$^\circ$ angle between the polarization directions of adjacent domains with a net in-plane polarization component $\mathbf{P}_\myparallel$ along $[100]_\textrm{pc}$; several additional films sharing the same domain pattern and wall density, with thicknesses from 20 to 100~nm, were also investigated. Mosaic-BFO (20\,nm thick) hosts a disordered domain pattern forming 71$^\circ$ angle between the corresponding domain polarizations exhibiting a net in-plane polarization component $\mathbf{P}_\myparallel$ along $[\bar{1}00]_\textrm{pc}$. The domain walls of the multidomain films sustain strong local electric fields, arising from the incompletely compensated bound charges~\cite{Yang2010,Guzelturk2020}. The films are excited at normal incidence through the substrate with 400~nm femtosecond pump pulses, above the band gap ($E_g\approx2.7$~eV~\cite{Sando2016b}). The light pump fluence was kept at $\approx$0.5~mJ/cm$^2$ for all the experiments unless specified. All angles are defined in the laboratory frame $(x,y,z)$ with $z$ along the surface normal: the pump light polarization angle $\alpha$ is measured from $x$, and $\varphi$ denotes the sample rotation about $z$ (Fig.~\ref{fig:fig2}\textbf{a}). The emitted THz field is detected by electro-optic sampling along $x$ or $y$, preserving both the amplitude and the phase of the transient (see Methods).

\subsection{Efficient bias-free THz emission from BiFeO$_3$ thin films}

First, we determine the THz emission of the four BiFeO$_3$ thin films and in a second step, we compare the BiFeO$_3$ THz emissions with spintronic FeCoB/Pt THz emitter. The THz electric fields emitted upon above-bandgap excitation from the four types of BiFeO$_3$ thin films are displayed in Fig.~\ref{fig:fig2}\textbf{b-e}. The THz transient of all four samples consists of an intense signal centered around time zero, accompanied by an oscillatory feature that persists for up to 4 picoseconds. In this work, we concentrate on the intense signal at around time zero as we are interested in the maximal THz emission. The most striking difference between the THz signals from the different samples is in their peak intensities. Oblique-BFO yields the lowest and Stripe-BFO the highest THz efficiency (see Supplementary Section~II for the THz transients for all the thicknesses). Intriguingly, not only the amplitudes changes but also the sign of the THz signal. Rotating the azimuthal angle $\varphi$ from $0^\circ$ (blue) to $180^\circ$ (red), results in an inversion of the THz transient polarity. This means that we can control the polarity of the emitted THz electric field simply by adjusting the sample orientation, without any external bias or electrical switching. These variations in both magnitude and sign hint at different physical origins for the THz emission process linked to ferroelectric domain orientations and patterns that will be scrutinized in section~\ref{sec:mecanisms}. The light pump fluence dependence of the peak THz electric field for Oblique-BFO, Out-of-plane-BFO and Stripe-BFO is shown in Fig.~\ref{fig:fig3}\textbf{a}. The THz field emitted by Oblique-BFO saturates (see Supplementary Section~III), while it remains linear for Out-of-plane-BFO and Stripe-BFO, the latter being the most efficient. The emitted THz field from Stripe-BFO saturates beyond a film thickness of $\sim$70~nm (Fig.~\ref{fig:fig3}\textbf{b}), consistent with the finite penetration depth of the 400~nm pump~\cite{Sando2018b}: the variation of the THz amplitude between films is well accounted for by the corresponding absorbed pump fluence (orange line in Fig.~\ref{fig:fig3}\textbf{b}). Mosaic-BFO, exhibiting a disordered domain pattern, yields a sizeable THz signal that falls on the same thickness dependence as the stripe-ordered films (Fig.~\ref{fig:fig3}\textbf{b}), its lower output relative to Stripe-BFO being fully accounted for by the reduced absorbed pump fluence. The frequency spectra of the four BiFeO$_3$ thin films, presented in Supplementary Section~IV, display similar features, limited primarily by the frequency response of our experimental setup. Upon tuning the pumping photon energy below the bandgap (800~nm), no signal is recorded for an equivalent fluence (see Supplementary Section~V). This indicates the importance of the generation of electron-hole pairs in the THz emission process. 

\begin{figure}[]
\includegraphics[width=\columnwidth]{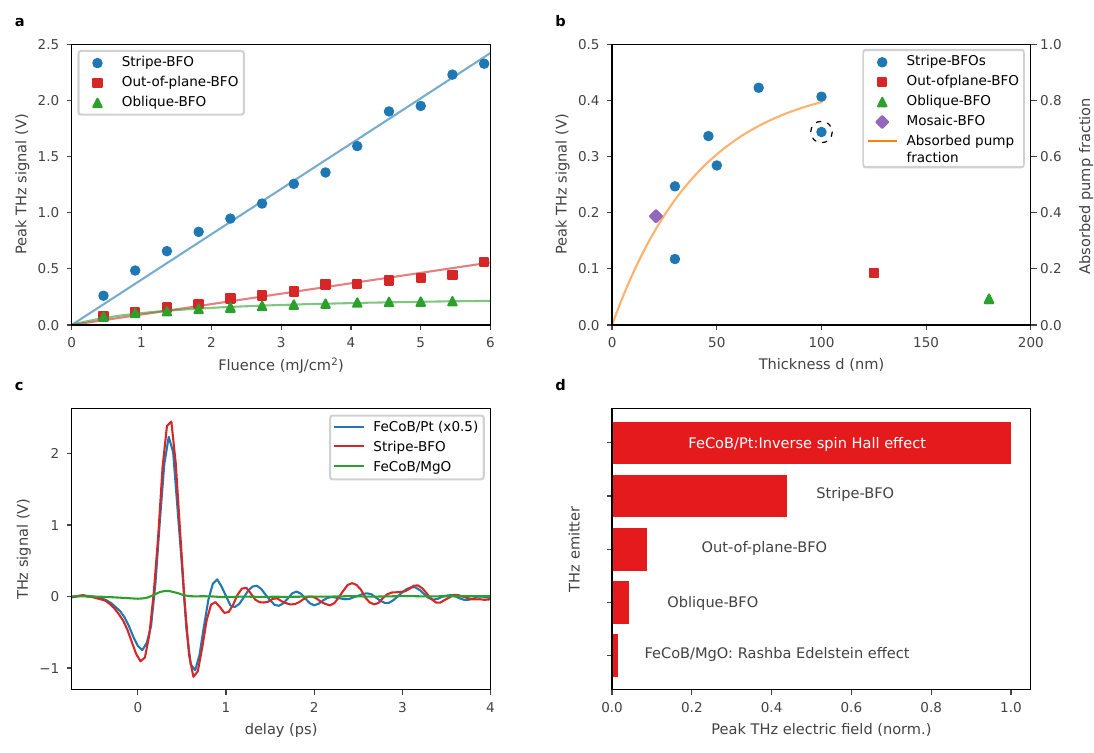}
\caption{\label{fig:fig3} \textbf{Comparison of THz emission from BiFeO$_3$ thin films and FeCoB/Pt spintronic emitters.} (a) Maximum amplitude of the THz field as a function of the light pump fluence for the BiFeO$_3$ thin films. (b) Thickness dependence of the peak THz electric field emitted by the BiFeO$_3$ thin films at a fluence of $\approx$0.5~mJ/cm$^2$. Dots: films with the stripe domain configuration (20--100~nm). The dashed circle marks the 100~nm film, referred to as Stripe-BFO, on which all angle- and time-resolved measurements are performed. Orange line and right axis: calculated absorbed pump fraction $A\propto1-e^{-d/\xi}$, with $\xi = $ 42 nm consistent with Ref.~\cite{Sando2018b}. (c) THz transients from Stripe-BFO and a spintronic emitter (FeCoB/Pt, scaled to 50\% of its maximum output) at their respective maximum emission capabilities. (d) THz electric field output from the BiFeO$_3$ thin films and spintronic emitters at their respective maximum emission capabilities, all normalized to the FeCoB/Pt bilayer response.}
\end{figure}

To contextualize our results, we compared the THz emission efficiency of the BiFeO$_3$ thin films to that of a well-established benchmark system, specifically the inverse spin Hall effect-based emitter FeCoB/Pt bilayer. In these spintronic emitters, an ultrafast optical pulse demagnetizes the ferromagnetic FeCoB layer, generating a spin current that is subsequently converted into a charge current via the inverse spin Hall effect in the adjacent Pt layer. This transient charge current causes the emission of a THz electric field (see, e.g.~\cite{Kampfrath2013,Seifert2017,Wu2021,Seifert2022}). The FeCoB(5nm)/Pt(3nm) spintronic emitter, previously characterized~\cite{Levchuk2023}, exhibits a standard response consistent with literature~\cite{seifert2016,Levchuk2023}. Remarkably, the maximum peak THz electric field emitted by Stripe-BFO reaches approximately half the maximum value observed for the FeCoB/Pt bilayer, as shown in Fig.~\ref{fig:fig3}\textbf{c} and~\textbf{d}. At these excitation densities the FeCoB/Pt emission is already saturated (see Supplementary Section~III), whereas Stripe-BFO remains in its linear regime, the maximum fluence being limited by our setup rather than by the sample. This leaves substantial headroom for the BiFeO$_3$ emitter, whose output may reach values comparable to FeCoB/Pt at higher excitation densities. Although the THz electric fields radiated from single-domain BiFeO$_3$ thin films are lower than those from multi-domain thin films with a macroscopic ferroelectric polarization, they remain substantial. Notably, these fields are at least an order of magnitude greater than those produced by emitters based on the inverse Edelstein effect, such as Rashba interfaces in FeCoB/MgO~\cite{Levchuk2023} and LaAlO$_3$/SrTiO$_3$~\cite{ElHamdi2024}. This result is summarized in the histograms of Fig.~\ref{fig:fig3}\textbf{d}. The substantial emitted THz signal together with the bias-free tunability of the THz polarization underline the potential of multifunctional BiFeO$_3$ thin films as robust THz sources for device applications.  

\subsection{Photocurrent contributions to the THz emission in BiFeO$_3$ thin films}
\label{sec:mecanisms}

Next, we identify the physical mechanisms giving rise to the THz emission. As below-bandgap excitation produces no detectable signal, non-resonant second-order nonlinearities such as optical rectification are excluded. The radiated THz field therefore originates from transient charge currents of photogenerated carriers~\cite{ShanJieandHeinz2004}. In non-centrosymmetric, ferroelectric, and antiferromagnetic materials such as BiFeO$_3$, three transient currents can arise in a photoexcited ferroelectric, each driven by a different aspect of the ferroelectric state  (see Methods). The most direct one involves the polarization itself: the photogenerated carriers screen the bound charges, transiently reducing $\mathbf{P}$, and the associated current 
\begin{equation}
\mathbf{J}^\textrm{P} = \partial\mathbf{P}/\partial t
\end{equation}
radiates~\cite{Takahashi2006,Rana2009,Levchuk2026} (see Methods). The second is driven by the electric fields that the polarization leaves uncompensated: the carriers drift in the internal field of the film, producing a photoconductive current $\mathbf{J}^\textrm{PC}$ and is expressed by: 
\begin{equation}
J^{\mathrm{PC}}_{i}=\sigma_{ijkl}E_{j}E^*_{k}E^{\mathrm{internal}}_{l}\ ,\label{eq:photo}
\end{equation} 
where $\sigma_{ijkl}$ is the photoconductive fourth-rank tensor. $E_j$ and $E^*_k$  are the electric field components of the incident light and its complex conjugation, respectively. $\mathbf{E}^{\mathrm{internal}}$ is the internal electric field arising from the ferroelectric polarization in the ferroelectric domain (if the depolarizing field is only partially screened~\cite{Ghosez2006}) or at domain-walls. The third does not require any pre-existing field but stems from the non-centrosymmetry of the lattice: interband excitation in a crystal lacking inversion symmetry generates a shift current, the bulk photovoltaic response $\mathbf{J}^\textrm{BPV}$~\cite{Sturman2020,Dai2023}. It  is expressed as:
\begin{equation}J^{\mathrm{BPV}}_{i}=\beta^\mathrm{BPV}_{ijk}E_{j}E^*_{k}\ ,\label{eq:bpv}\end{equation}
where $\beta^\mathrm{BPV}_{ijk}$ is the third-rank bulk photovoltaic tensor. Magnetization-related contributions are negligible under our experimental conditions: in particular, the light-induced transient magnetization, governed in BiFeO$_3$ by the spin cycloid~\cite{Fishman2018}, averages to zero, as detailed in Supplementary Section~VI. Consequently, the source term is restricted to light-induced charge currents, and the emitted THz electric field from a thin-film emitter follows $\mathbf{E}_\textrm{THz} \propto \mathbf{J}^\textrm{total}$~\cite{ShanJieandHeinz2004,Guzelturk2020,Seifert2022}, with
\begin{equation}
\mathbf{J}^\textrm{total} =
\underbrace{(\textrm{polarization screening})}_{\mathbf{J}^\mathrm{P}}
+ \underbrace{(\textrm{drift in internal field})}_{\mathbf{J}^\mathrm{PC}}
+ \underbrace{(\textrm{bulk photovoltaic effect})}_{\mathbf{J}^\mathrm{BPV}}.
\label{eq:Jtotal}
\end{equation}

As these three currents add coherently in the emitted field, their contributions cannot be separated in a single measurement. We therefore disentangle them through their distinct angular fingerprints in $\alpha$ and $\varphi$ (see Methods). $\mathbf{J}^\textrm{P}$ is a vector locked to the ferroelectric polarization: independent of the pump polarization, its in-plane projection simply rotates with the sample, contributing an isotropic (constant) term in $\alpha$ and a $\cos\varphi$ component in $\varphi$. $\mathbf{J}^\textrm{BPV}$ and $\mathbf{J}^\textrm{PC}$ are instead tensorial responses, evaluated throughout using the symmetry-allowed elements of the rhombohedral $R3c$ phase of bulk BiFeO$_3$ (point group $3m$). As the photoconductive current driven by the component of $\mathbf{E}^{\mathrm{internal}}$ along $\mathbf{P}$ shares the $R3c$ angular form of the bulk photovoltaic current, we group them as $\mathbf{J}^\textrm{Ani.} = \mathbf{J}^\textrm{BPV} + \mathbf{J}^\textrm{PC}$. Depending on the film orientation, $\mathbf{J}^\textrm{Ani.}$ contributes an isotropic term and a $\cos2\alpha$ component in $\alpha$, together with $\cos\varphi$ and $\cos3\varphi$ components in $\varphi$, as summarized in Table~\ref{tab:harmonics} (see Methods for the derivations). The $\cos2\alpha$ and $\cos3\varphi$ components are thus exclusive signatures of $\mathbf{J}^\textrm{Ani.}$, whereas the isotropic and $\cos\varphi$ components can receive contributions from both $\mathbf{J}^\mathrm{P}$ and $\mathbf{J}^\textrm{Ani.}$. These fingerprints are moreover robust against the direction of the internal field: any component of $\mathbf{E}^\textrm{internal}$ perpendicular to $\mathbf{P}$ generates only harmonics already contained in Table~\ref{tab:harmonics}, modifying their relative weights but not the allowed sets (see Methods). Finally, the weak anisotropy of the optical absorption introduces a negligible pump-polarization dependence of the photocarrier density (Supplementary Section~VII). 

\begin{table}[htb]
\centering
\caption{Theoretical angular dependence of the in-plane photocurrents in the laboratory frame for the four crystalline orientations, for the tensorial current $\mathbf{J}^\textrm{Ani.}$, evaluated with the tensor elements of the rhombohedral $R3c$ structure of BiFeO$_3$ (point group $3m$), and the screening current $\mathbf{J}^\textrm{P}$. The entries give the angular harmonics of the emitted THz field as a function of the pump polarization angle $\alpha$ at fixed $\varphi$, and of the sample azimuthal angle $\varphi$ at fixed $\alpha$: $n\alpha$ and $m\varphi$ denote components varying as $\cos(n\alpha)$ and $\cos(m\varphi)$, and ``iso'' the angle-independent term. For Stripe-BFO and Mosaic-BFO, the response is averaged over the two ferroelastic variants. See Supplementary Section~VIII for polar-plot illustrations of these harmonics.}
\label{tab:harmonics}
\begin{tabular}{lcccc}
\hline\hline
 & \multicolumn{2}{c}{$\mathbf{J}^\textrm{Ani.}$} & \multicolumn{2}{c}{$\mathbf{J}^\textrm{P}$} \\
\cline{2-3}\cline{4-5}
Sample & $\alpha$-scan & $\varphi$-scan & $\alpha$-scan & $\varphi$-scan \\
\hline
Oblique-BFO       & iso $+\, 2\alpha$    & $1\varphi + 3\varphi$  & iso         & $1\varphi$  \\
Out-of-plane-BFO& $2\alpha$            & $3\varphi$             & Absent$^{a}$ & Absent$^{a}$ \\
Stripe-BFO        & iso $+\, 2\alpha$    & $1\varphi + 3\varphi$  & iso         & $1\varphi$  \\
Mosaic-BFO        & iso $+\, 2\alpha$    & $1\varphi + 3\varphi$  & iso         & $1\varphi$  \\
\hline\hline
\end{tabular}\\[2pt]
\raggedright\footnotesize $^{a}$\,$\mathbf{J}^\textrm{P}$ is directed along the surface normal and does not radiate into the laboratory $x-y$ plane.
\end{table}

\begin{figure}[!ht]
\includegraphics[width=\columnwidth]{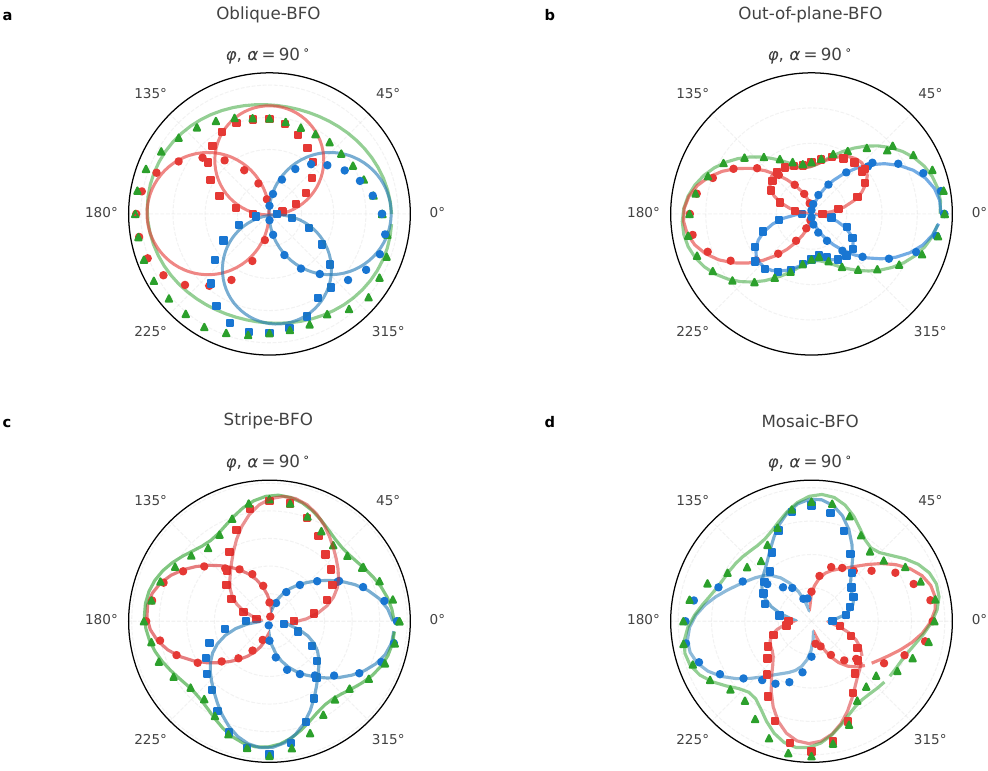}
\caption{\label{fig:fig4} \textbf{Angle-resolved THz emission from BiFeO$_3$ thin films: $\varphi$-scan.} Sample azimuthal angle ($\varphi$) dependence of the peak THz field amplitude in the time domain: $x$-detection (dots) and $y$-detection (squares), color-encoded by sign (red: positive; blue: negative), and total field amplitude $E_\textrm{tot}(\varphi)=\sqrt{E_x^2 + E_y^2}$ (green triangles). Each panel is a polar plot for corresponding to Oblique-BFO, Out-of-plane-BFO, Stripe-BFO, and Mosaic-BFO. See Supplementary Section~VIII for illustrative examples of the individual angular harmonics in Cartesian and polar representations. Raw data uncorrected for the substrate effect on the pump pulse. See Methods and Supplementary Section~XIII. Lines: corrected harmonic decomposition with the substrate-induced pulse splitting reapplied (Methods).}
\end{figure}

\subsubsection{Amplitude of the emitted THz field}

We now confront these theoretical predictions with the measured angular dependence, starting with the peak THz field in the time domain data. Figure~\ref{fig:fig4} shows the peak emitted field along $x$ (dots) and $y$ (squares) as a function of $\varphi$ at fixed $\alpha=90^\circ$. Oblique-BFO, Stripe-BFO, and Mosaic-BFO display a strong two-lobed signature rotating by $90^\circ$ between the two detection channels, the fingerprint of a dominant $1\varphi$ contribution ($\cos\varphi$ in $E_x$, $\sin\varphi$ in $E_y$), which can arise from $\mathbf{J}^\textrm{P}$ and/or $\mathbf{J}^\textrm{Ani.}$. The nearly undistorted lobes of Oblique-BFO point to a weak tensorial contribution (Supplementary Section~IX). For the multidomain samples the lobes are distorted, partly by the birefringence of the DyScO$_3$ substrate, which splits the pump pulse into two delayed components and is corrected for in the spectral analysis below (Methods and Supplementary Section~XIII). Their tensorial content is therefore assessed there. Out-of-plane-BFO differs qualitatively, with a distorted two-lobed pattern along $x$ and an asymmetric four-lobed pattern along $y$, indicating coexisting $\cos\varphi$ and $\cos3\varphi$ contributions (Supplementary Section~IX). Finally, the polarity of the emitted field reflects the direction of the photocurrent: for all films with a non-zero $\mathbf{P}_\myparallel$, the photocurrent is globally antiparallel to $\mathbf{P}_\myparallel$. Combining the two detection channels into the total in-plane field amplitude $E_\textrm{tot}$ (green triangles) removes the detection-axis dependence: a circular $E_\textrm{tot}(\varphi)$ indicates a single dominant angular component, whereas deviations from circularity reveal interference between coexisting contributions or an anisotropy of the response (see Methods). The nearly circular $E_\textrm{tot}$ of Oblique-BFO is consistent with a dominant $\mathbf{J}^\textrm{P}$. Stripe-BFO and Mosaic-BFO show maxima at $\varphi = 0$, $90$, $180$, and $270^\circ$, i.e.\ when the pump polarization lies along a substrate axis, as expected from the pulse splitting. Out-of-plane-BFO is the most anisotropic, although its ideal response predicts a circle, already signaling the symmetry-forbidden components and the in-plane anisotropy discussed below. See Supplementary Section~IX for additional measurements.

A quantitative separation of the coexisting currents requires the complex spectral representation of the emitted field: the relative weights and phases of the harmonic components, and the correction of the pump-pulse splitting induced by the substrate birefringence (Methods and Supplementary Section~XIII), are both defined in the frequency domain. In Fig.~\ref{fig:fig5}\textbf{a--d}, we display the corrected angular dependence of the emitted field amplitude on $\alpha$ at 1.0~THz (Supplementary Section~IX for additional data). After correction, the patterns are frequency-independent within our band, indicating that a single current family dominates the response of each film. Oblique-BFO, Stripe-BFO, and Mosaic-BFO are isotropic, whereas Out-of-plane-BFO shows a two-lobed shape with a non-zero minimum, revealing a sizeable isotropic offset on top of the $\cos2\alpha$ component, although the ideal $R3c$ response of this orientation predicts a pure $\cos2\alpha$ dependence with no in-plane $\mathbf{J}^\textrm{P}$ contribution (Table~\ref{tab:harmonics}). The amplitude alone, however, discards both the sign of the underlying photocurrent and the relative temporal phase between coexisting contributions (Supplementary Section~VIII). Both are contained in the full complex spectrum of the transients, whose amplitude and phase are directly measured by electro-optic sampling, and which we exploit next.

\subsubsection{Angular harmonics of the complex THz field}

\begin{figure}[!ht]
\includegraphics[width=\columnwidth]{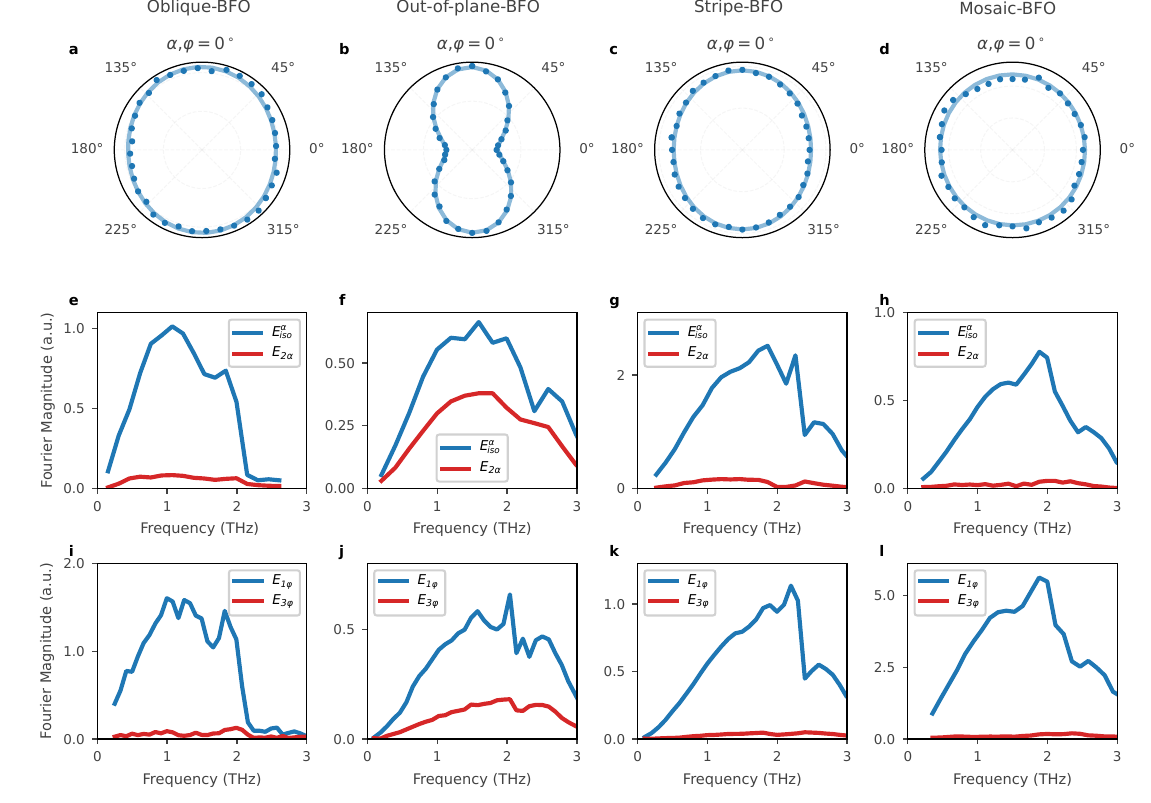}
\caption{\label{fig:fig5} \textbf{Distinct photocurrents underlying THz emission in BiFeO$_3$ thin films.} Columns correspond to Oblique-BFO, Out-of-plane-BFO, Stripe-BFO and Mosaic-BFO. (a--d)~Angular dependence of the emitted field amplitude at 1.0~THz on the pump polarization angle $\alpha$, corrected for the substrate-induced pump-pulse splitting (Methods); lines are the harmonic decomposition into isotropic and $\cos2\alpha$ components. The patterns are frequency-independent within our band (Supplementary Section~X). (e--h)~Magnitude of the symmetry-allowed $\alpha$-harmonics $\tilde{E}^\alpha_\textrm{iso}$ (blue) and $\tilde{E}_{2\alpha}$ (red). (i--l)~Magnitude of the symmetry-allowed azimuthal harmonics $\tilde{E}_{1\varphi}$ (blue) and $\tilde{E}_{3\varphi}$ (red). All spectra are corrected for the substrate-induced pump-pulse splitting (Methods). The reconstructed time-domain channel transients are shown in Supplementary Section~XII.}
\end{figure}

In the following, we therefore decompose the complex spectra into angular harmonics of the pump polarization angle $\alpha$, $\tilde{E}(\omega,\alpha) = \tilde{E}^\alpha_\textrm{iso}(\omega) + \tilde{E}_{2\alpha}(\omega)\cos (2\alpha)$, and of the azimuthal angle $\varphi$, $\tilde{E}(\omega,\varphi) = \tilde{E}_{1\varphi}(\omega)\cos (\varphi) + \tilde{E}_{3\varphi}(\omega)\cos (3\varphi)$, whose complex coefficients give access to the spectral weight and the relative phase of each underlying current (Fig.~\ref{fig:fig5}\textbf{e--l}; see Methods). In all samples, the extracted harmonics are restricted to precisely these sets: all other harmonics, forbidden by symmetry, remain below roughly 10\% of the dominant contribution over the whole spectral range (see Supplementary Section~XII). As the coefficients carry the full spectral amplitude and phase, they can be transformed back to the time domain: the reconstructed transients display the actual temporal profile of the field radiated by each angular component (Supplementary Section~XII). Coexisting photocurrents have previously been disentangled in THz emission through the symmetry of the transients at topological-insulator surfaces~\cite{Braun2016}, and through their distinct temporal structure in semiconductors~\cite{Laman2005} and ferroelectric SbSI~\cite{Sotome2019}. Here, we combine both aspects in a frequency-resolved complex harmonic analysis, giving direct access to the relative spectral phase of each angular component. To identify the underlying currents, we compare, for each sample, the spectral weights and relative phases of these components with the fingerprints of Table~\ref{tab:harmonics}. The lines in Figs.~\ref{fig:fig4} and~\ref{fig:fig5}\textbf{a--d} show the result of the harmonic decomposition, reproducing the angular patterns.

We begin with Oblique-BFO. Over the whole spectral range, $\left|\tilde{E}^\alpha_\textrm{iso}(\omega)\right|$ dominates over $\left|\tilde{E}_{2\alpha}(\omega)\right|$ (Fig.~\ref{fig:fig5}\textbf{a} and~\textbf{e}), and  only the $1\varphi$ harmonic is present with $3\varphi$ harmonic negligible (Fig.~\ref{fig:fig5}\textbf{i}). This is the fingerprint of $\mathbf{J}^\textrm{P}$ in Table~\ref{tab:harmonics}: a pump-polarization-independent current attached to the crystal, whose in-plane projection follows the sample rotation and the absence of the $2\alpha$ and $3\varphi$ harmonics shows a negligible $\mathbf{J}^\textrm{Ani.}$. The antiparallel photocurrent direction established above is precisely the signature expected for a transient reduction of the polarization. Together with the saturation of the emitted field at high pump fluence (Fig.~\ref{fig:fig3}\textbf{a} and Supplementary Section~III), these observations identify the ultrafast screening of $\mathbf{P}$ by photocarriers as the dominant emission mechanism in Oblique-BFO~\cite{Rana2009}. This assignment is in agreement with time-resolved second-harmonic-generation imaging of BiFeO$_3$ thin films, which directly revealed a substantial sub-picosecond quench of the ferroelectric polarization upon femtosecond excitation~\cite{Levchuk2026}.

Out-of-plane-BFO shows sizeable isotropic, $2\alpha$, $1\varphi$, and $3\varphi$ contributions, also evidenced by the strongly distorted $\varphi$-scan (Fig.~\ref{fig:fig4}\textbf{b}). The $2\alpha$ and $3\varphi$ harmonics are the signature of $\mathbf{J}^\textrm{Ani.}$. With the ferroelectric polarization along the surface normal, however, $\mathbf{J}^\textrm{P}$ does not radiate into the detection direction and can account for neither the isotropic nor the $1\varphi$ component. A slight triclinic distortion, as reported by Dufour~\textit{et al.}~\cite{Dufour2023}, relaxes the $R3c$ constraints and, through modified tensor elements and/or an in-plane internal field, accounts for the observed isotropic and $1\varphi$ components (see Methods). Indeed, the lobes of the $1\varphi$ component are oriented along the $[\bar1\bar12]_\textrm{pc}$ axis identified by X-ray diffraction~\cite{Dufour2023}, locking these anomalous components to the structural in-plane anisotropic strain of the film. Consistently, the reconstructed THz transients oscillate in phase (Supplementary Section~XII), indicating that a single current type dominates. The fluence dependence identifies which one: the emission grows linearly up to our highest excitation densities (Fig.~\ref{fig:fig3}\textbf{a}), whereas $\mathbf{J}^\textrm{PC}$ would saturate once the internal field is screened. As such, we identify $\mathbf{J}^\textrm{BPV}$ as the dominant source of the emission in Out-of-plane-BFO.

For the multidomain samples Stripe-BFO and Mosaic-BFO, the corrected response is isotropic in $\alpha$ and $1\varphi$ in azimuth (Fig.~\ref{fig:fig5}\textbf{c,d,g,h,k,l}). As the two films contain the same two ferroelastic variants, they share the same net in-plane polarization and domain-averaged tensorial response: their nearly identical angular content confirms that it is set by the ferroelectric variant populations rather than by the stripe or mosaic arrangement. This response is that of the screening current $\mathbf{J}^\textrm{P}$: its polarity, along $[\bar{1}00]_\textrm{pc}$ (Stripe-BFO) or $[100]_\textrm{pc}$ (Mosaic-BFO), is antiparallel to $\mathbf{P}_\parallel$ and agrees with that reported for stripe-domain BiFeO$_3$ under identical above-bandgap excitation, there attributed to charge separation in the built-in domain-wall fields~\cite{Guzelturk2020}. This assignment is, moreover, independent of the substrate correction, which affects only the tensorial channels (see Methods). The domain-wall picture and the screening of $\mathbf{P}_\parallel$ describe, in fact, the same process: the wall fields stem from the incompletely compensated bound charge, and the charge separation they drive is the microscopic pathway of the screening. A photoconductive current driven by fixed built-in fields would in addition carry a $\cos2\alpha$ component on these multidomains films (Table~\ref{tab:harmonics}), which is not observed. Rather than adding a channel with a distinct angular signature, the multidomain geometry thus enhances the screening current, through the larger reservoir of uncompensated bound charge, the higher photocarrier density at the walls~\cite{Bhatnagar2013}, and the efficient charge separation in their built-in fields~\cite{Yang2010}, consistent with the higher output and unsaturated fluence dependence of Stripe-BFO (Fig.~\ref{fig:fig3}\textbf{a}).

Taken together, the three domain configurations favor distinct dominant photocurrents: the screening current $\mathbf{J}^\textrm{P}$ in Oblique-BFO, the bulk photovoltaic current $\mathbf{J}^\textrm{BPV}$ in Out-of-plane-BFO, and, in the multidomain films, a wall-enhanced screening current. The ferroelectric domain configuration thus governs not only the emission strength but also the balance between the underlying photocurrents.

\section{Summary}

We have investigated the THz electric field radiated upon ultrashort optical excitation from four epitaxial BiFeO$_3$ thin films with distinct ferroelectric configurations: an oblique polarization (Oblique-BFO), an out-of-plane polarization (Out-of-plane-BFO), and two multidomain configurations (Stripe-BFO and Mosaic-BFO). Our results demonstrate that (i) the THz emission from BiFeO$_3$ thin films is comparable in magnitude to that of the spintronic emitter FeCoB/Pt, and (ii) the amplitude of the radiated field is effectively controlled by the ferroelectric domain configuration. Notably, Stripe-BFO surpasses the other configurations of comparable thickness by a factor of four, up to the highest fluences explored. Through the angular harmonic decomposition of the complex THz field as a function of pump polarization, azimuthal angle, and fluence, we assign this enhanced response to a dominant ultrafast screening of the ferroelectric polarization, assisted by the strong local electric fields of the multidomain configuration. Our findings establish the ferroelectric domain configuration as a design parameter for light-driven currents: it sets not only the strength of the THz emission but also the microscopic mechanism that generates it. Ferroelectric emitters thereby combine bias-free, magnet-free operation with an emitted field whose polarity encodes the nonvolatile ferroelectric polarization, and whose strength and origin are set by the domain configuration.

\section*{Acknowledgments}
We acknowledge support from the French Agence Nationale de la Recherche (ANR) through the EPHONO (ANR-22-CE30-0007), THz-Mufins (ANR-21-CE42-0030), RELAXTRON (ANR-23-CE24-0011), DEMIETTES (ANR25-CE24-5503) and SAFER (ANR-25-CE09-0835) projects. This work is supported by a France 2030 government grant managed by the French National Research Agency (ANR-22-PEEL-0010).

\section*{Methods}

\subsection{BiFeO$_3$ thin films}
The four types of epitaxial BiFeO$_3$ films with distinct ferroelectric domain configuration are the following. For details on growth and characterization see Supplementary Section~I. We introduce an orthonormal frame $(X, Y, Z)$ attached to the crystallographic axes of each sample (see Fig.~\ref{fig:fig1}\textbf{a} for the correspondence with the pseudocubic directions), with $X$ and $Y$ parallel to the sample edges and $Z$ parallel to $z$, i.e. perpendicular to the sample surface. In the following, the indices c, pc, h, and o denote the cubic, pseudocubic, hexagonal, and orthorhombic settings, respectively. \\
\textbf{Oblique-BFO:} BiFeO$_3$(110)$_\mathrm{pc}$/SrTiO$_3$(110)$_\mathrm{c}$ (180\,nm), single-domain, with its polarization oblique to the surface: the out-of-plane component $\mathbf{P}_{\perp}$ points toward the substrate, and the in-plane component $\mathbf{P}_\myparallel$ lies along $[00\bar{1}]_\mathrm{pc}$. \\
\textbf{Out-of-plane-BFO:} BiFeO$_3$(111)$_\mathrm{pc}$/DyScO$_3$(011)$_\mathrm{o}$ (125\,nm), single-domain, with the polarization perpendicular to the surface and pointing toward the substrate; X-ray diffraction further indicates the presence of an anisotropic in-plane strain in the BiFeO$_3$ thin film along $[\bar{1}\bar{1}2]_\textrm{pc}$~\cite{Dufour2023}. \\
\textbf{Stripe-BFO:} BiFeO$_3$(001)$_\mathrm{pc}$/DyScO$_3$(110)$_\mathrm{o}$ (100\,nm) exhibits a periodic stripe domain pattern (9\,DW\,µm$^{-1}$) with 71$^\circ$ angle between the polarization directions of adjacent domains. Its net macroscopic polarization has an out-of-plane component $\mathbf{P}_{\perp}$ toward the substrate, and an in-plane component $\mathbf{P}_\myparallel$ along $[100]_\textrm{pc}$. Several additional films sharing the same domain pattern and wall density, with thicknesses from 20 to 100\,nm, were also investigated. \\
\textbf{Mosaic-BFO:} BiFeO$_3$(001)$_\mathrm{pc}$/DyScO$_3$(110)$_\mathrm{o}$ (20\,nm) hosts a disordered domain pattern forming 71$^\circ$ angle between the corresponding domain polarizations. The film retains a macroscopic net polarization, with an out-of-plane component $\mathbf{P}_{\perp}$ toward the substrate, and $\mathbf{P}_\myparallel$ along $[\bar{1}00]_\textrm{pc}$. The domain walls of the multidomain films sustain strong local electric fields, arising from the incompletely compensated bound charge~\cite{Yang2010,Guzelturk2020}. 

\subsection{X-ray diffraction}
The structure of the films was characterized by X-ray diffraction using a PANalytical (Empyrean) X-ray diffractometer and a Rigaku SmartLab diffractometer, both equipped with a Ge(220)×2 monochromator selecting Cu-K$\alpha_1$ radiation in parallel beam configuration.

\subsection{Piezoresponse force microscopy}
The PFM images were acquired with an atomic force microscope (Nanoscope V multimode, Bruker) coupled with two external lock-in amplifiers (SR830, Stanford Research) to record simultaneously the vertical and lateral responses. An AC voltage oscillating at a frequency of 35 kHz was applied between the SrRuO$_3$ or La$_{0.7}$Sr$_{0.3}$MnO$_3$ bottom electrode and a grounded Pt-coated tip. We used cantilevers with stiffnesses ranging from 7 to 40 N.m$^{-1}$. 




\subsection{THz emission spectroscopy}

Experiments were carried out with an amplified laser of \SI{35}{fs} at 1 kHz, operating at the center wavelength of 800 nm. The output laser beam was split, where most of the laser power was frequency-doubled in a non-linear crystal, then used to excite the BiFeO$_3$ thin films above their bandgap, while the remaining power was used for electro-optic detection~\cite{Planken2001}. The unfocused ultrashort laser pulse propagates along the $z$-direction and enters the sample through the substrate side. This configuration prevents any spurious effects that could arise from THz field propagation through the substrate. Normal incidence also avoids Fresnel contributions, which are challenging to account for in optically birefringent materials such as BiFeO$_3$ (see Guzelturk et al.~\cite{Guzelturk2020}). The effect of the substrate on the pump laser pulse are taken into account (see below). The emitted THz field is analyzed by electro-optic sampling in a 0.5~mm thick $\langle 110 \rangle$ ZnTe crystal. The detection bandwidth is limited from 0.2 to roughly 3\, THz. Electro-optic detection measures the THz electric field and is therefore phase-sensitive, preserving both the amplitude and the phase of the emitted transient~\cite{seifert2016}. The horizontal ($x$) or vertical ($y$) component of the THz field is selected by rotating the ZnTe crystal by $90^\circ$ between the two configurations; the experiment is thereby sensitive exclusively to the in-plane current dynamics within the laboratory $x$--$y$ plane. The probe power on the photodiodes was kept constant for all measurements, and the photodiode output signal was recorded by lock-in detection, yielding the THz field in volts. All the THz optical path from the emitter to the electro-optic detection crystal was located in a dry air-purged sealed box to avoid water absorption during the THz beam propagation. The half-wave plate (for adjusting the pump polarization angle $\alpha$) and the sample (for azimuthal angle $\varphi$) are mounted on motorized rotation stages, enabling automated acquisition of the full THz transients over evenly spaced angles covering 360$^\circ$ in both $\alpha$ and $\varphi$: the completeness and regularity of the angular sampling ensure an unbiased extraction of the harmonic coefficients.  

\subsection{Total in-plane field amplitude}
The total in-plane field amplitude $E_\textrm{tot}(\varphi) = \sqrt{E_x^2 + E_y^2}$ is independent of the detection axis. A circular $E_\textrm{tot}(\varphi)$, i.e. a field of constant magnitude that only rotates with the sample, indicates a single dominant angular component, whereas deviations from circularity reveal the interference between coexisting contributions. A single angular component of the current, a fixed in-plane vector rotating with the sample, or the pure $\cos(2\alpha - 3\varphi)$ response of the ideal $(111)_\textrm{pc}$ BiFeO$_3$ thin film orientation, whose two in-plane projections form a quadrature pair of equal amplitude, yields a constant $|J|$ and hence a circular $E_\textrm{tot}(\varphi)$. When two components of different angular order coexist, their cross term modulates $E_\textrm{tot}$. For coexisting $\cos\varphi$ and $\cos3\varphi$ contributions, the beating produces a $4\varphi$ modulation, whose depth reflects their amplitude ratio.


\subsection{Ultrafast screening of the ferroelectric polarization in BiFeO$_3$}
The photogenerated carriers redistribute on a sub-picosecond timescale and partially screen the bound charge of the ferroelectric polarization, transiently reducing its effective value~\cite{Takahashi2006,Sotome2019,Levchuk2026}. The resulting time-dependent polarization $\mathbf{J}^{\mathrm{P}}\propto\partial \mathbf{P}/\partial t$ radiates in the THz range. We assume that the photoinduced change in the ferroelectric polarization remains collinear with the equilibrium ferroelectric polarization, i.e. the screening transiently reduces its magnitude without rotating its direction at short timescales~\cite{Levchuk2026}, remaining collinear with the equilibrium polarization along $[111]_\textrm{pc}$. For this assumption, $\mathbf{J}^\textrm{P}$ is expressed as:
\begin{equation}\label{eq:partialp}
\mathbf{J}^\textrm{P}=\frac{\partial \mathbf{P}}{\partial t}=\frac{\partial \left|\mathbf{P}\right|}{\partial t}\, \mathbf{e}_{P}
\end{equation}
with $\mathbf{e}_{P}$ the unit vector along $\mathbf{P}$.
If the screening transiently reduces its magnitude without rotating its direction, $\partial_t\left|\mathbf{P}\right|<0$ and thus $\textbf{J}^\textrm{P}$ is pointing to $-\mathbf{e}_{P}$, antiparallel to $\mathbf{P}$. The angular dependence of $\mathbf{J}^\textrm{P}$ follows directly from expressing $\mathbf{e}_P$ in the laboratory frame for each crystalline orientation, averaged over domain variants for Stripe-BFO/Mosaic-BFO. As $\mathbf{J}^\textrm{P}$ is attached to the crystal and independent of the pump polarization, its in-plane projection is $\alpha$-independent and exhibits a pure $1\varphi$ azimuthal dependence for all the samples but Out-of-plane-BFO. Indeed, for Out-of-plane-BFO, $\textbf{J}^\textrm{P}$ reads in the laboratory $(x,y,z)$ frame:
\begin{equation}
\textbf{J}^\textrm{P}(\varphi) \propto I_\textrm{Inc.}^2
\left(
\begin{array}{c}
0\\
0\\
1\\
\end{array}
\right).
\end{equation}
For this crystalline structure, we cannot detect this effect as it does not radiate in the $x-y$ laboratory plane. For Oblique-BFO, it reads:
\begin{equation}
\textbf{J}^\textrm{P}(\varphi) \propto \dfrac{I_\textrm{Inc.}^2}{\sqrt{3}} 
\left(
\begin{array}{c}
 -\cos (\varphi )\\
 -\sin (\varphi )
  \\
\sqrt{2} \\
\end{array}
\right),
\end{equation}
As for the Stripes/Mosaic-BFO, Eq.~\ref{eq:partialp} holds for each variant separately. the average current is then expressed as:
\begin{equation}
\textbf{J}^\textrm{P}(\varphi) \propto \dfrac{I_\textrm{Inc.}^2}{\sqrt{3}} 
\left(
\begin{array}{c}
 -\cos (\varphi )\\
 -\sin (\varphi )
  \\
1\\
\end{array}
\right),
\end{equation}

\subsection{Bulk photo-voltaic effect in BiFeO$_3$}
The phenomenological expression of the bulk photo-voltaic current $\mathbf{J}^\textrm{BPV}$ generation upon excitation of a non-centrosymmetric material by light radiation is given by~\cite{Sturman2020,Dai2023}:
\begin{equation}
    J^\textrm{BPV}_i=\beta^\mathrm{BPV}_{ijk}E_jE^\star_k,
\end{equation}
where $\beta^\mathrm{BPV}$ is the third-rank bulk photovoltaic tensor. $E_j$ and $E^*_k$  are the electric field components of the incident light and its complex conjugation, respectively. In the laboratory frame, the expression of the pump light electric field propagating along the $z$-axis is given by $\mathbf{E}_{\textrm{pump}}=E_\textrm{Inc.}\exp\left(i(kz-\omega t)\right)\cos\left(\alpha\right) \mathbf{e_x}+E_\textrm{Inc.}\exp\left(i(kz-\omega t)\right)\sin\left(\alpha\right)\mathbf{e_y}$. 

Under the assumption that the structural distortions are minor in the films, it is justified to use a quasi-rhombohedral approximation and the R3c point group. Upon symmetry operation, the BPV tensor for the R3c point group is expressed by, with index 3 referring to the polar axis:
\begin{equation}\beta^\mathrm{BPV}=
\left(\begin{array}{@{}ccc|ccc|ccc@{}}
0 & -\beta _{222} & \beta _{113} & -\beta _{222} & 0 & 0 & \beta _{311} & 0 & 0\\
 -\beta _{222} & 0 & 0 &0 & \beta _{222} & \beta _{113} & 0 & \beta _{311} & 0  \\
 \beta _{131} & 0 & 0 & 0 & \beta _{131} & 0 & 0 & 0 & \beta _{333} \\ 
\end{array}\right).
\end{equation}

The hexagonal unit cell axis is usually defined in pseudocubic coordinates by:
\begin{equation}
\left\{
\begin{aligned}\nonumber
x'_h     &=\dfrac{1}{\sqrt{2}} [1\bar{1}0]_{pc} \\
y'_h    &=\dfrac{1}{\sqrt{6}} [11\bar{2}]_{pc} \\
z'_h   &=\dfrac{1}{\sqrt{3}} [111]_{pc} \\
\end{aligned}
\right.
\end{equation}
where $[001]_h=[111]_{pc}$ is the direction of the ferroelectric polarization.

For the crystallographic orientation of Out-Of-Plane-BFO, one can express the BPV current $\boldsymbol{J^\textrm{BPV}}$ as a function of the angle $\alpha$ by in the laboratory frame $(x,y,z)$:
\begin{equation}\label{eq:bpvout}
\textbf{J}^\textrm{BPV}(\alpha,\varphi)=\left(
\begin{array}{c}
 J^\textrm{BPV}_x \\
 J^\textrm{BPV}_y \\
 J^\textrm{BPV}_z \\
\end{array}
\right)=E_\textrm{Inc.}^2
\left(
\begin{array}{c}
 \beta_{222} \cos (2 \alpha -3 \varphi ) \\
 -\beta_{222} \sin (2 \alpha -3 \varphi ) 
  \\
 \beta_{311} \\
\end{array}
\right).
\end{equation}
The expression of $\boldsymbol{J^\textrm{BPV}}$ is rather simple as the crystallographic orientation of the thin film is epitaxially aligned with the underlying unit cell, such that the film’s lattice axes coincide with those of the crystal structure. We recover the $3m$ symmetry signature with the $3\varphi$ period upon sample rotation as well as the $2\alpha$ dependence for the pump light polarization. This fact is not true for Oblique-BFO, Stripes-BFO, and Mosaic-BFO, hence the more complicated expression for $\boldsymbol{J^\textrm{BPV}}$. The full expressions of $\boldsymbol{J^\textrm{BPV}}$ are given in Supplementary Section~X. Here, we only provide the expressions in the relevant $x-y$ laboratory frame for our detection geometry.

In the case of Oblique-BFO the BPV current $\boldsymbol{J^\textrm{BPV}}$ as a function of the angles $\alpha$ and $\varphi$ in the laboratory $(x,y)$ plane is:
\begin{equation}\label{eq:bpvout2}
\textbf{J}^\textrm{BPV}(\alpha,\varphi)\propto E_\textrm{Inc.}^2
\left(
\begin{array}{c}
A\cos (2 \alpha -3 \varphi )+B \cos (2 \alpha -\varphi ) + C \cos (\varphi )\\
 -A\sin (2 \alpha -3 \varphi )-B\sin (2 \alpha -\varphi ) +C \sin (\varphi )
 \\
\end{array}
\right).
\end{equation}
$A$, $B$, and $C$ are constants given by combinations of the tensor elements $\beta_{113}$, $\beta_{222}$, $\beta_{311}$, and $\beta_{333}$ (see Supplementary Section~X for their explicit expressions). We find that $\mathbf{J}^\textrm{BPV}$ depends on both angles, contributing $\cos\varphi$ and $\cos3\varphi$ components in $\varphi$ and, at fixed azimuth, an isotropic term and a $\cos2\alpha$ modulation in $\alpha$.

Stripes-BFO and Mosaic-BFO are organized as a stripe domain pattern or disordered pattern, respectively. As such, following the work of Young~\textit{et al} and Bhatnagar~\textit{et al}~\cite{Young2012,Bhatnagar2013}, the BPV current is calculated as the average value for the two ferroelastic variants BPV response. $\boldsymbol{J^\textrm{BPV}}$ is expressed, a function of angles $\alpha$ and $\varphi$ in the laboratory frame $(x,y)$ by:
\begin{align}\label{eq:bpvmulti2}
&\textbf{J}^\textrm{BPV}(\alpha,\varphi)\propto E_\textrm{Inc.}^2\times\\
&\left(
\begin{array}{c}
 D \cos (2 \alpha -3 \varphi ) +E\cos (2 \alpha -\varphi )+\sqrt{2} \beta_{222} (5 \sin (2 \alpha -3 \varphi )-\sin (2 \alpha -\varphi )+2 \sin (\varphi ))+F \cos (\varphi ) \\
 -D \sin (2 \alpha -3 \varphi )+E\sin (2 \alpha -\varphi )+\sqrt{2} \beta_{222} (5 \cos (2 \alpha -3 \varphi )+\cos (2 \alpha -\varphi )-2 \cos (\varphi ))+F \sin (\varphi )
\end{array}
\right).\nonumber
\end{align}
$D$, $E$, and $F$ are constants given by combinations of the tensor elements $\beta_{113}$, $\beta_{222}$, $\beta_{311}$, and $\beta_{333}$ (see Supplementary Section~X for their explicit expressions). We find that $\mathbf{J}^\textrm{BPV}$ depends on both angles, contributing $\cos\varphi$ and $\cos3\varphi$ components in $\varphi$ and, at fixed azimuth, an isotropic term and a $\cos2\alpha$ modulation in $\alpha$.

\subsection{Photoconductive effect in BiFeO$_3$}

Photoexcited carriers accelerated by the static internal electric field associated with the ferroelectric polarization, produce a transient drift current $\mathbf{J}^{\mathrm{PC}}$. It is expressed by: 
\begin{equation}
J^{\mathrm{PC}}_{i}=\sigma_{ijkl}E_{j}E^*_{k}E^{\mathrm{internal}}_{l}\ ,\label{eq:photo}
\end{equation} 
where $\sigma_{ijkl}$ is the photoconductive fourth-rank tensor. $E_j$ and $E^*_k$  are the electric field components of the incident light and its complex conjugation, respectively. $\mathbf{E}^{\mathrm{internal}}$ is the internal electric field arising from the ferroelectric polarization in the ferroelectric domain (if the depolarizing field is only partially screened~\cite{Ghosez2006}) or at domain-walls. 

It is worth noting the role of the internal electric field direction for $\mathbf{J}^\textrm{PC}$ in the symmetry of the response. A perfectly screened single-domain crystal has no $\mathbf{E}^\textrm{internal}$, so that $\mathbf{J}^\textrm{PC}$ vanishes. In BiFeO$_3$ thin films, however, the bound charge at interfaces and domain walls is only partially compensated by free carriers, charged defects, and electrodes, leaving a residual depolarizing field that acts on the photocarriers. 
The direction of $\mathbf{E}^{\mathrm{internal}}$ within a domain is not known a priori: depending on the electrostatic boundary conditions, it may deviate from the polarization direction~\cite{Ghosez2006}. We therefore decompose it into components parallel and perpendicular to the polarization,
\begin{equation}
\mathbf{E}^{\mathrm{internal}} = E_{\parallel}\,\mathbf{e}_P + \mathbf{E}_{\perp},
\end{equation}
which, Eq.~\eqref{eq:photo} being linear in $\mathbf{E}^{\mathrm{internal}}$, splits the photoconductive current exactly into two terms, $\mathbf{J}^{\mathrm{PC}} = \mathbf{J}^{\mathrm{PC}}_{\parallel} + \mathbf{J}^{\mathrm{PC}}_{\perp}$. The parallel term reduces the fourth-rank response to an effective third-rank tensor, $\sigma^{\mathrm{eff}}_{ijk} = E_{\parallel}\,\sigma_{ijk3}$, which possesses the same $R3c$ symmetry-allowed structure as the bulk photovoltaic tensor $\beta^\mathrm{BPV}$: $\mathbf{J}^{\mathrm{PC}}_{\parallel}$ is therefore angularly indistinguishable from $\mathbf{J}^{\mathrm{BPV}}$, with effective elements $\sigma^{\mathrm{eff}}_{222} = -E_{\parallel}\sigma_{2223}$, $\sigma^{\mathrm{eff}}_{311} = E_{\parallel}\sigma_{3113}$, [\ldots] (see below for the derivation).  The perpendicular term $\mathbf{E}_{\perp}$ lowers the effective symmetry of the response, but generates no angular components outside the sets of Table~\ref{tab:harmonics} (see below for the derivation). Moreover, in the multidomain films, the residual depolarizing field of each domain is supplemented by the strong local fields at the domain walls, arising from their incompletely compensated bound charge~\cite{Yang2010,Guzelturk2020}. Directed along the in-plane wall normal, these wall fields contribute through both projections: their component along each variant's polarization adds to $\mathbf{J}^{\mathrm{PC}}_{\parallel}$, while the perpendicular part populates only the shared isotropic and $\cos\varphi$ channels and adds to $\mathbf{J}^{\mathrm{PC}}_{\perp}$. Together with the equivalence of $\mathbf{J}^{\mathrm{PC}}_{\parallel}$ and $\mathbf{J}^{\mathrm{BPV}}$ established above, this motivates the grouping $\mathbf{J}^{\mathrm{Ani.}} = \mathbf{J}^{\mathrm{BPV}} + \mathbf{J}^{\mathrm{PC}}$ used in the main text.



If $\mathbf{E}^\textrm{internal}$ is collinear with $\mathbf{P}$, $\textbf{J}^\textrm{PC}$ can be described with a reduced rank-3 tensor $\sigma^\mathrm{Eff.}$ knowing that $E^\mathrm{internal}$ is fixed along the $[001]_h$ direction in the hexagonal unit cell of BiFeO$_3$:
\begin{equation}
    J^\textrm{PC}_i=\left |E_0^\mathrm{internal}\right|\sigma^\mathrm{Eff.}_{ijk}E_j(\omega)E^\star_k(\omega) \label{eq:photo3},
\end{equation}
with $E_0^{\mathrm{internal}}$ the depolarizing electric field amplitude and the rank-3 tensor $\sigma^\mathrm{Eff.}$ expressed by:
\begin{equation}\sigma^\mathrm{Eff.}=
\left(\begin{array}{@{}ccc|ccc|ccc@{}}
0 & -\sigma _{2223} & \sigma _{1133} & -\sigma _{2223} & 0 & 0 & \sigma _{3113} & 0 & 0\\
 -\sigma _{2223} & 0 & 0 &0 & \sigma _{2223} & \sigma _{1113} & 0 & \sigma _{3113} & 0  \\
 \sigma _{1113} & 0 & 0 & 0 &\sigma _{3113} & 0 & 0 & 0 & \sigma _{3333} \\ 
\end{array}\right),
\end{equation}

It can be noted that the tensors $\sigma^{\mathrm{Eff.}}_{ijk}$ and $\beta^{\mathrm{BPV}}_{ijk}$ share the same symmetry elements. Consequently, given the non-zero tensor elements of $\sigma^{\mathrm{Eff.}}$, the angular dependence of $\mathbf{J}^{\mathrm{PC}}(\alpha,\varphi)$ is \emph{a priori} identical to that of $\mathbf{J}^{\mathrm{BPV}}(\alpha,\varphi)$. 

For Out-of-Plane-BFO with an internal electric field along [111]$_\textrm{pc}$, $\textbf{J}^\textrm{PC}(\alpha,\varphi)$ is expressed by:

\begin{equation}
\textbf{J}^\textrm{PC}(\alpha,\varphi)=\left(
\begin{array}{c}
 J^\textrm{PC}_x \\
 J^\textrm{PC}_y \\
 J^\textrm{PC}_z \\
\end{array}
\right)=\left|E_0^\mathrm{internal}\right| E_\textrm{Inc.}^2
\left(
\begin{array}{c}
 \sigma_{2223} \cos (2 \alpha -3 \varphi ) \\
 -\sigma_{2223} \sin (2 \alpha -3 \varphi ) 
  \\
 \sigma_{3113} \\
\end{array}
\right).
\end{equation}
One can note that the $3\varphi$ and $2\alpha$ harmonics are possible, equivalently to $\mathbf{J}^\textrm{BPV}$ defined in Eq.~\ref{eq:bpvout}.
However, if $\mathbf{E}^\textrm{internal}$ departs from this ideal case, as shown in the following, lower harmonics will be possible. If the internal electric field is along $X=[\bar{1}\bar{1}2]_{pc}$, $\textbf{J}^\textrm{PC}(\alpha,\varphi)$ reads:
\begin{align}
&\textbf{J}^\textrm{PC}(\alpha,\varphi)=\left(
\begin{array}{c}
 J^\textrm{PC}_x \\
 J^\textrm{PC}_y \\
 J^\textrm{PC}_z \\
\end{array}
\right)=\left|E_0^\mathrm{internal}\right| E_\textrm{Inc.}^2\times
\left(
\begin{array}{c}
  \sigma_{1122} \cos (2 \alpha -\varphi )+(\sigma_{1111}-\sigma_{1122}) \cos (\varphi ) \\
 \sigma_{1122} \sin (2 \alpha -\varphi )+(\sigma_{1111}-\sigma_{1122}) \sin (\varphi )
  \\
 \sigma_{3112} \cos (2 (\alpha -\varphi ))\\
\end{array}
\right).
\end{align}
If the internal electric field is along $Y=[1\bar{1}0]_{pc}$, $\textbf{J}^\textrm{PC}(\alpha,\varphi)$ reads:
\begin{align}
&\textbf{J}^\textrm{PC}(\alpha,\varphi)=\left(
\begin{array}{c}
 J^\textrm{PC}_x \\
 J^\textrm{PC}_y \\
 J^\textrm{PC}_z \\
\end{array}
\right)=\left|E_0^\mathrm{internal}\right| E_\textrm{Inc.}^2\times
\left(
\begin{array}{c}
 \sigma_{1122} \sin (2 \alpha -\varphi )+(\sigma_{1122}-\sigma_{1111}) \sin (\varphi )\\
 -\sigma_{1122} \cos (2 \alpha -\varphi )+  (\sigma_{1111}-\sigma_{1122}) \cos (\varphi ) \\
 -\sigma_{3112} \sin (2 (\alpha -\varphi ))
\\
\end{array}
\right).
\end{align}
For these electric field orientations, the $3\varphi$ harmonic is lost and the $1\varphi$ and $2\alpha$ harmonics are possible in the detection $x-y$ plane. 

The same reasoning applies to Oblique-BFO and the multidomain films: any component of $\mathbf{E}^\textrm{internal}$ perpendicular to $\mathbf{P}$ generates only harmonics that are already allowed for $\mathbf{J}^\textrm{BPV}$ at these crystalline orientations (Eqs.~\ref{eq:bpvout2} and~\ref{eq:bpvmulti2}) and listed in Table~\ref{tab:harmonics}. The grouping $\mathbf{J}^{\mathrm{Ani.}} = \mathbf{J}^{\mathrm{BPV}} + \mathbf{J}^{\mathrm{PC}}$ therefore holds for all samples. See Supplementary Section~XI for the complete expression.

\subsection{Substrate birefringence and walk-off correction}

The DyScO$_3$ substrates are birefringent at the pump wavelength. A \SI{400}{nm} pulse polarized at an angle $\alpha-\psi$ to the substrate fast axis (orientation $\psi$) is split into two orthogonally polarized components, carrying the fractions $\cos^2(\alpha-\psi)$ and $\sin^2(\alpha-\psi)$ of the pump energy along the fast and slow axes, which reach the film with a group delay $\Delta t$ of \SI{132}{fs} and \SI{80}{fs} for the (110)$_\mathrm{o}$ and (011)$_\mathrm{o}$ cuts, respectively (measured in situ from the emission timing and independently by spectral interferometry through the bare substrates, Supplementary Section~XIII). As $\Delta t$ exceeds the pump duration, the film is excited by two successive linearly polarized pulses, along the fast axis and then along the slow axis, and never experiences an intermediate polarization state; rotating $\alpha$ merely redistributes the pump energy between the two. No splitting occurs on the cubic SrTiO$_3$ substrate of Oblique-BFO.

The measured spectrum is therefore the weighted sum of the spectra $\tilde{E}_\mathrm{f}$ and $\tilde{E}_\mathrm{s}$ emitted under fast- and slow-axis pumping, the latter delayed:
\begin{equation}
  \tilde{E}(\omega;\alpha)
  = \cos^{2}(\alpha-\psi)\,\tilde{E}_\mathrm{f}
  + \sin^{2}(\alpha-\psi)\,z\,\tilde{E}_\mathrm{s}
  = \underbrace{\tfrac12[\tilde{E}_\mathrm{f} + z\tilde{E}_\mathrm{s}]}_{\tilde{E}^{\alpha}_{\mathrm{iso}}}
  + \underbrace{\tfrac12[\tilde{E}_\mathrm{f} - z\tilde{E}_\mathrm{s}]}_{\tilde{E}_{2\alpha}}\cos2(\alpha-\psi),
  \quad z = e^{-i\omega\Delta t}.
\end{equation}
Three consequences follow. (i)~Even for a film responding identically to both pump polarizations ($\tilde{E}_\mathrm{s} = \tilde{E}_\mathrm{f}$), the uncorrected data show a spurious $\cos2\alpha$ component; conversely a purely tensorial response leaks into the isotropic channel by the same factor. (ii)~The $\sin2(\alpha-\psi)$ quadrature, which requires both field components simultaneously, is not excited and is not reported. (iii)~In azimuthal scans the substrate co-rotates with the film, $\psi = \psi_0 + \varphi$, and the varying weights generate spurious odd harmonics; no azimuthal information is lost.

Pump-polarization-independent currents are unaffected by mechanism: the screening current is set by the absorbed energy, not by the polarization, so both pulses generate the same emission ($\tilde{E}_\mathrm{s} = \tilde{E}_\mathrm{f}$) and its fingerprint, isotropic in $\alpha$, $\cos\varphi$ in azimuth, with its polarity, is invariant under the splitting at any $\Delta t$: the artifact lives in the difference of the two emitted spectra, the screening current in their sum.

For each detected THz polarization component ($E_x$ or $E_y$), the spectra of a scan are decomposed per frequency in the substrate frame, $\tilde{E} = A + B\cos2(\alpha-\psi) + C\sin2(\alpha-\psi)$ ($C$ must vanish and serves as a null test); the two emitted spectra are recovered, $\tilde{E}_\mathrm{f} = A+B$, $\tilde{E}_\mathrm{s} = (A-B)\,e^{+i\omega\Delta t}$, and recombined without the delay, $\tilde{E}^{\mathrm{corr}}_{\mathrm{iso},2\alpha} = \tfrac12(\tilde{E}_\mathrm{f} \pm \tilde{E}_\mathrm{s})$. Azimuthal scans are corrected with the same $(\Delta t,\psi_0)$. An error in $(\Delta t,\psi_0)$ leaves residual spurious quadrature and can only overestimate tensorial contributions. All quantities in Fig.~5 and in the Supplementary Information are walk-off corrected; Fig.~4 shows raw data. Validation is given in Supplementary Section~XIII.

\subsection{Harmonic decomposition in the pump polarization angle}
The polar plots in Fig.~\ref{fig:fig5} display the amplitude of the emitted THz field and therefore discard the phase information carried by the electro-optic detection. Because each current source radiates with its own temporal dynamics, it contributes to the complex THz spectrum with a distinct spectral phase; the phase is also what the walk-off correction (previous section) operates on. We therefore fit the complex spectrum, rather than its modulus, with a harmonic decomposition in the pump polarization angle $\alpha$,
\begin{equation}
\tilde{E}(\omega,\alpha) = \tilde{E}^\alpha_\textrm{iso}(\omega) + \tilde{E}_{2\alpha}(\omega)\cos(2\alpha - \alpha_0),
\end{equation}
or in the sample azimuthal angle $\varphi$,
\begin{equation}
\tilde{E}(\omega,\varphi) = \tilde{E}_{1\varphi}(\omega)\cos(\varphi - \varphi_1) + \tilde{E}_{3\varphi}(\omega)\cos(3\varphi - \varphi_3),
\end{equation}
where $\tilde{E}^\alpha_\textrm{iso}(\omega)$, $\tilde{E}_{2\alpha}(\omega)$, $\tilde{E}_{1\varphi}(\omega)$, and $\tilde{E}_{3\varphi}(\omega)$ are complex coefficients whose amplitudes give the spectral weight of the contributions, and whose relative phase $\Delta\phi^\alpha(\omega) = \arg\tilde{E}_{2\alpha} - \arg\tilde{E}^\alpha_\textrm{iso}$ or $\Delta\phi^\varphi(\omega) = \arg\tilde{E}_{3\varphi} - \arg\tilde{E}_{1\varphi}$ tests whether the underlying currents share a common temporal profile. On the DyScO$_3$-mounted films the coefficients are walk-off corrected before analysis (previous section); on SrTiO$_3$ the phase $\alpha_0$ is free. An example of such decomposition for FeCoB(5\,nm)/Pt(3\,nm) is displayed in Supplementary Section~XII: the harmonic contributions ($1\alpha$, $2\alpha$, \dots) are negligible compared to the isotropic one, as expected for the emission mechanism of spintronic emitters.

\subsection{Relative phase of the harmonic components} 

For films much thinner than the THz wavelength, the emitted field follows the sheet current, $E_\textrm{THz}(t) \propto J(t)$~\cite{ShanJieandHeinz2004}. A current following the pump intensity envelope and a current following the photocarrier population, $J(t) \propto \int^t I(t')\,dt'$, radiate fields whose spectra differ by a factor $1/(i\omega)$: they are in quadrature at all frequencies, with a spectral-weight ratio increasing with frequency. Harmonic components sharing a common temporal profile are therefore in phase, whereas a quadrature between them signals distinct dynamics, a criterion previously used to separate coexisting photocurrents~\cite{Laman2005,Sotome2019}, and the signature that a substrate-induced pulse splitting mimics (previous section).

\bibliography{export2}

\end{document}


\preprint{APS/123-QED}

\title{Supplementary Informations for \\ Domain-engineered ferroelectric BiFeO$_3$ thin films for efficient bias-free THz emission}

\author{Z. Abdul Hadi$^{1}$, N. Rezi$^{2}$, M. Monti$^{1}$, G. Vaudel$^{1}$,  A. Abdelsamie$^{2}$, C. Carrétéro$^{2}$, D. Sando$^{3,4}$, M. Viret$^{5}$, V. Garcia$^{2}$, S. Fusil$^{2}$, M. C. Weber$^{1}$, P. Ruello$^{1}$, V. Juv\'e $^{1}$\footnote{Electronic address: vincent.juve@univ-lemans.fr}}

\affiliation{
$^{1}$ Le Mans Université, CNRS, IMMM, Institut des Molécules et des Matériaux du Mans, UMR 6283, F-72000 Le Mans, France\\
$^{2}$ Laboratoire Albert Fert, CNRS, Thales, Universit\'e Paris-Saclay, 91767 Palaiseau, France\\
$^{3}$ The MacDiarmid Institute for Advanced Materials and Nanotechnology, School of Physical \& Chemical Sciences, University of Canterbury, New Zealand\\
$^{4}$ School of Materials Science and Engineering, UNSW Sydney, Australia\\
$^{5}$ SPEC UMR CEA/CNRS, Université Paris-Saclay, L’Orme les Merisiers, 91191 Cedex, Gif-sur-Yvette, France\\
}




\date{\today}

\maketitle

\section{Thin film epitaxial growth, structural characterization and ferroelectric domain configuration.}

\subsection{Oblique-BFO: Single-domain BiFeO$_3$(110)$_{pc}$ thin film}
\begin{figure}[!h]
\includegraphics[width=\columnwidth]{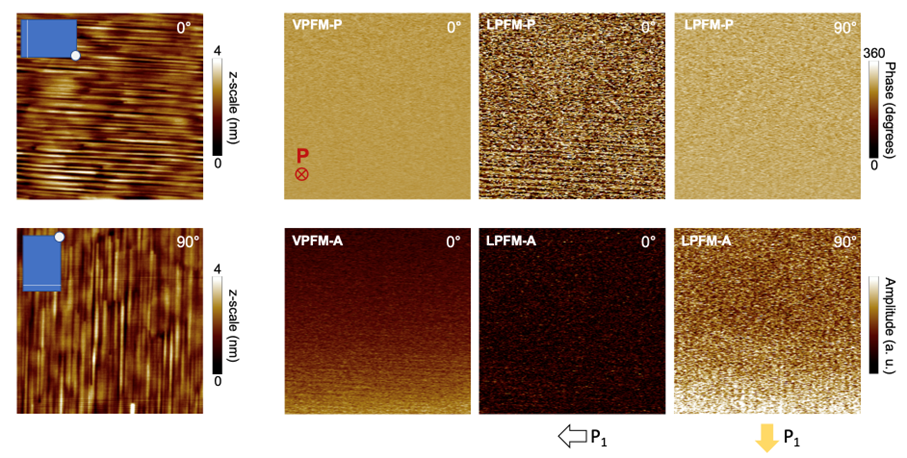}
\caption{PFM characterization of the as-grown state of BiFeO$_3$/SrRuO$_3$/SrTiO$_3$(110)$_c$. The vertical-PFM phase (VPFM-P) and amplitude (VPFM-A) indicate a uniform polarization pointing downward. The lateral-PFM phase (LPFM-P) shows noise and the amplitude (LPFM-A) virtually no signal at 0$^\circ$. In contrast, the LPFM-P is uniform and bright at 90$^\circ$, with a well-defined LPFM-A. These results suggest a single domain polarization state. In all the PFM images, the cantilever is horizontal and pointing to the left.}\label{fig:laf1}
\end{figure}

A single-domain epitaxial BiFeO$_3$(110)$_{pc}$ thin film was grown by pulsed laser deposition on a SrRuO$_3$-buffered SrTiO$_3$(110)$_{c}$ substrate (as defined in Refs.~\cite{Sando2019,Gu2023}. Ablation was carried out by a KrF laser with wavelength $\lambda$ = 248 nm at 5 Hz. The 12-nm layer of SrRuO$_3$ was grown from a stoichiometric target at 660$^\circ$C in 100 mtorr of oxygen with fluence of $\approx$2 J/cm2. The BiFeO$_3$ layer was grown from a Bi$_{1.1}$FeO$_3$ ceramic target, while the substrate was held at 590$^\circ$C in a background oxygen pressure of 100 mtorr. The thickness of the BiFeO$_3$ layer is estimated at 180$\pm$10 nm, extrapolated from the growth rate, taken from a sample grown directly beforehand on which x-ray reflectivity calibration was performed. PFM characterization was carried out (Figure~S\ref{fig:laf1}). Vertical PFM phase and amplitude indicate a homogeneous downward polarization state. Lateral PFM at 0° shows no signal and a uniform bright phase with a well-defined amplitude at 90°, confirming the single ferroelectric domain nature of the thin film with a tilted polarization state. Considering these experiments, we are able to assign precisely the ferroelectric polarization with respect to the sample directions for THz emission experiments.

\subsection{Out-of-plane-BFO: Single-domain BiFeO$_3$(111)$_{pc}$ thin film}
\begin{figure}[!h]
\includegraphics[width=0.5\columnwidth]{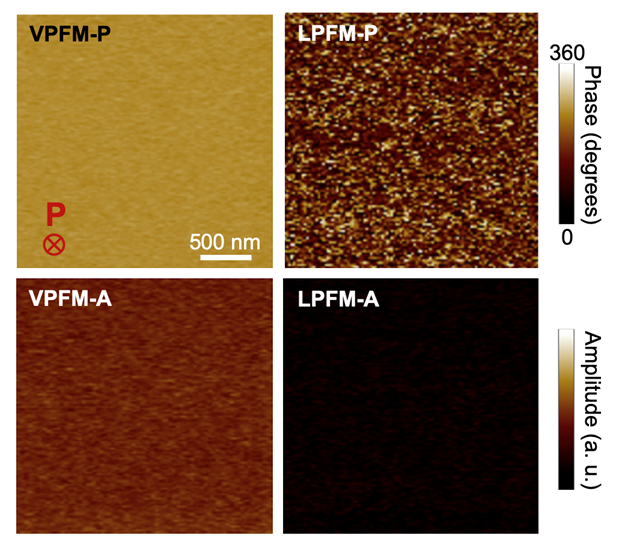}
\caption{PFM characterization of the as-grown state of BiFeO$_3$/SrRuO$_3$/DyScO$_3$(011)$_o$. The vertical-PFM phase (VPFM-P) and amplitude (VPFM-A) indicate a uniform polarization pointing downward. The lateral-PFM phase (LPFM-P) shows noise and the amplitude (LPFM-A) virtually no signal, in line with a purely vertical polarization direction. In all the PFM images, the cantilever is horizontal and pointing to the left.}\label{fig:laf2}
\end{figure}

The epitaxial thin film of BiFeO$_3$(111)$_{pc}$ was grown on a DyScO$_3$(011)$_o$ substrate by pulsed laser deposition using a KrF excimer laser (as defined in~\cite{Dufour2023}). The scandate substrate was preliminary annealed under constant oxygen flow at 1000$^\circ$C for 3 hours. A 12-nm-thick bottom electrode of SrRuO$_3$ was first grown under 0.2 mbar of oxygen at 660$^\circ$C. The 125-nm-thick BiFeO$_3$ thin film was grown under 0.36 mbar of oxygen at a temperature of 670$^\circ$C. The whole heterostructure was grown with a laser repetition rate of 5 Hz. The film thickness was estimated from X-ray reflectivity on thinner samples and double-checked by performing C-AFM experiments while indenting the film. PFM experiments reveal a uniform vertical PFM signal and no lateral PFM response (Figure~S\ref{fig:laf2}), indicating that the film polarization is purely vertical and pointing toward the SrRuO$_3$ electrode.

\begin{figure}[!h]
\includegraphics[width=\columnwidth]{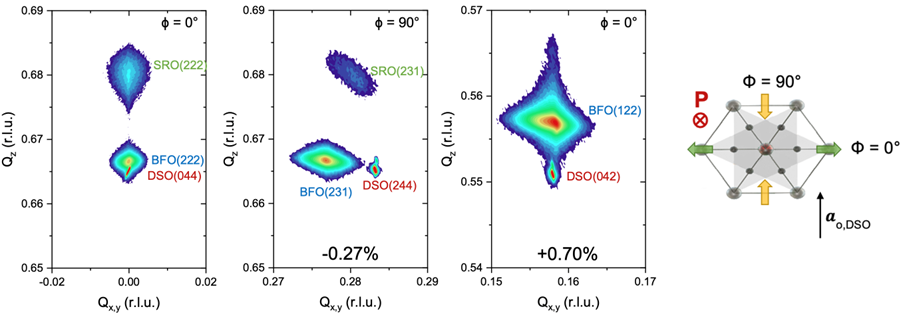}
\caption{X-ray diffraction reciprocal space maps around the (044)$_o$, (244)$_o$ and (042)$_o$ peaks of DyScO$_3$. The BiFeO$_3$ film is fully strained ($\phi = 0^\circ$) along DyScO$_3$(042)$_o$, and we estimate an epitaxial tensile strain of +0.70$\%$ for this in-plane direction. In contrast, the film is not fully strained along the a$_o$ axis of DyScO$_3$ ($\phi = 90^\circ$), and the resulting compressive strain is -0.27$\%$. As a result, the hexagonal (111)$_{pc}$ plane undergoes an anisotropic distortion, as emphasized by the sketch (as defined in~\cite{Dufour2023}).}\label{fig:laf3}
\end{figure}

X-ray diffraction characterization (Figure~S\ref{fig:laf3}) further confirms the single-domain nature of the (111)$_{pc}$ film, and reveal that the orthorhombic DyScO$_3$ substrate induces some anisotropic in-plane strain in the BiFeO$_3$ film~\cite{Dufour2023}. Hence, we can consider that the resulting symmetry of the film is no longer rhombohedral but rather triclinic.

\subsection{Stripe-BFO: Stripe-domain BiFeO$_3$(001)$_{pc}$ thin film}
\begin{figure}[!h]
\includegraphics[width=\columnwidth]{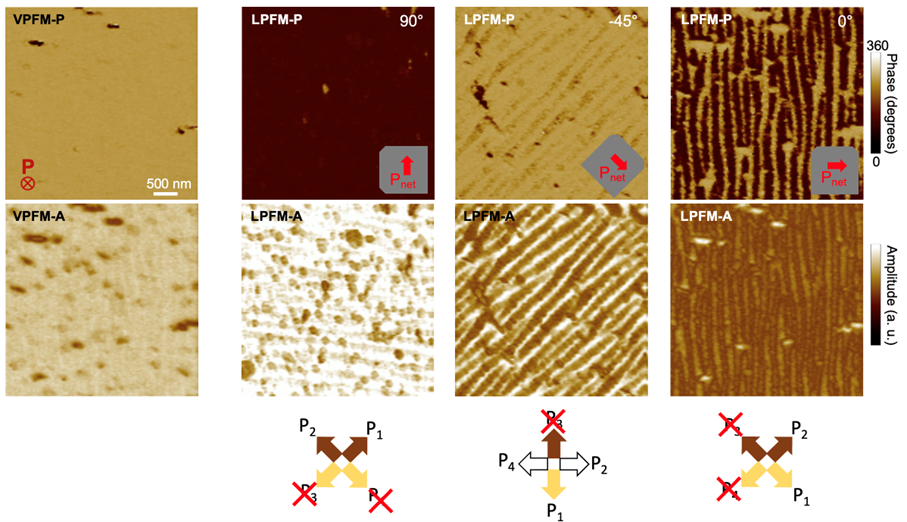}
\caption{PFM characterization of the as-grown state of BiFeO$_3$/La$_{0.7}$Sr$_{0.3}$MnO$_3$/DyScO$_3$(110)$_o$. The vertical-PFM phase (VPFM-P) and amplitude (VPFM-A) indicate a uniform polarization pointing downward. The lateral-PFM phase (LPFM-P) and amplitude (LPFM-A) are collected under different angles (90$^\circ$, -45$^\circ$, 0$^\circ$) of the PFM cantilever with respect to the sample to discard the possible in-plane variants of BiFeO$_3$ (P$_1$, P$_2$, P$_3$, P$_4$). The LPFM-P image at 0$^\circ$ shows a self-organized stripe-domain pattern, with two alternating ferroelectric variants. For an angle of 90$^\circ$, the LPFM-P becomes uniform and dark, indicating that only the P$_1$ and P$_2$ variants are present. This is confirmed at -45$^\circ$, where, out of the two possible P$_3$ and P$_1$ variants, only the P$_1$ variant is detected in the LPFM-P image while the P$_2$ variant along the cantilever cannot be probed and shows a zero LPFM-A signal. Consequently, vector PFM imaging enables us to know that P$_1$ and P$_2$ are present in the film, with 71$^\circ$ domain walls and the corresponding net in-plane polarization $P_{net}$, as displayed in the sketch (red arrow). In all the PFM images, the cantilever is horizontal and pointing to the left.}\label{fig:laf4}
\end{figure}

The epitaxial thin film of BiFeO$_3$(001)$_{pc}$ was grown by pulsed laser deposition using a KrF excimer laser on a DyScO$_3$(110)$_o$ single crystal substrate (as defined in~\cite{Gross2017}). The La$_{0.7}$Sr$_{0.3}$MnO$_3$ bottom electrode (1.2 nm) was grown with 5 Hz repetition rate at 660$^\circ$ under 0.2 mbar of oxygen. The BiFeO$_3$ film (120 nm) was subsequently grown at the same temperature under 0.36 mbar of oxygen with 1 Hz repetition rate. The sample was slowly cooled down under high oxygen pressure. PFM characterization (Figure~S\ref{fig:laf4}) reveals a homogeneous downward polarization (VPFM phase and amplitude) with two ferroelectric variants self-organized in the form of stripes (LPFM-P at 0$^\circ$). Vector PFM was performed in order to detect the possible ferroelectric variants that are present in the stripes. Lateral PFM at 90$^\circ$ indicates a uniform dark phase signal suggesting that only P$_1$ and P$_2$ are present (see the sketch). This is confirmed by lateral PFM at -45$^\circ$ where only one of the two variants (P$_1$) is giving piezoresponse, as the other (P$_2$) is aligned parallel to the cantilever. Hence, the self-organized stripe-domain pattern corresponds to alternating P$_1$ and P$_2$ variants with 71$^\circ$ domain walls.

\begin{figure}[!h]
\includegraphics[width=\columnwidth]{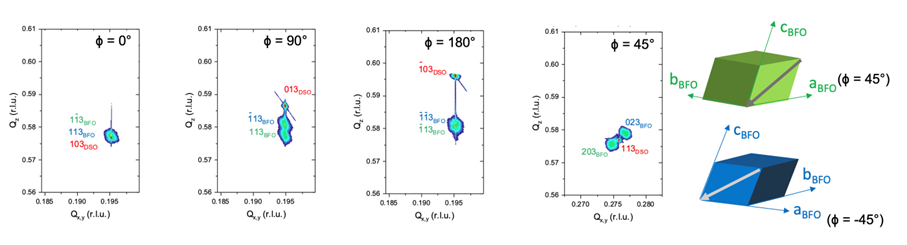}
\caption{X-ray diffraction reciprocal space maps around the (103)$_m$, (013)$_m$ (-103)$_m$ and (113)$_m$ peaks of DyScO$_3$. The BiFeO$_3$ film is fully strained ($\phi=0^\circ$,$90^\circ$,$180^\circ$) by the DyScO$_3$ substrate, as emphasized by the same in-plane Q$_{x,y}$ values between DyScO$_3$ and BiFeO$_3$. Two ferroelastic monoclinic variants of BiFeO$_3$ can be distinguished as emphasized for $\phi=45^\circ$ and the vertical Q$_z$ splitting for $\phi=90^\circ$. These results are in agreement with a stripe domain sample of alternating ferroelastic monoclinic variants, in which the polarization is rotating by 90$^\circ$ in the film plane (see the sketch).}\label{fig:laf5}
\end{figure}

The corresponding ferroelastic variants associated to these ferroelectric domains were detected using X-ray diffraction reciprocal space maps (Figure~S\ref{fig:laf5}). Reciprocal space maps at $\phi=0^\circ$, 90$^\circ$, 180$^\circ$ show that the BiFeO$_3$ film is fully strained by the DyScO$_3$ substrate. Maps at $\phi=90^\circ$ and 45$^\circ$ show two peaks corresponding to the two monoclinic variants of BiFeO$_3$ arranged with 71$^\circ$  domain walls (see the sketch).

\subsection{Mosaic-BFO: Disordered striped BiFeO$_3$(001)$_{pc}$ thin film}
\begin{figure}[!h]
\includegraphics[width=\columnwidth]{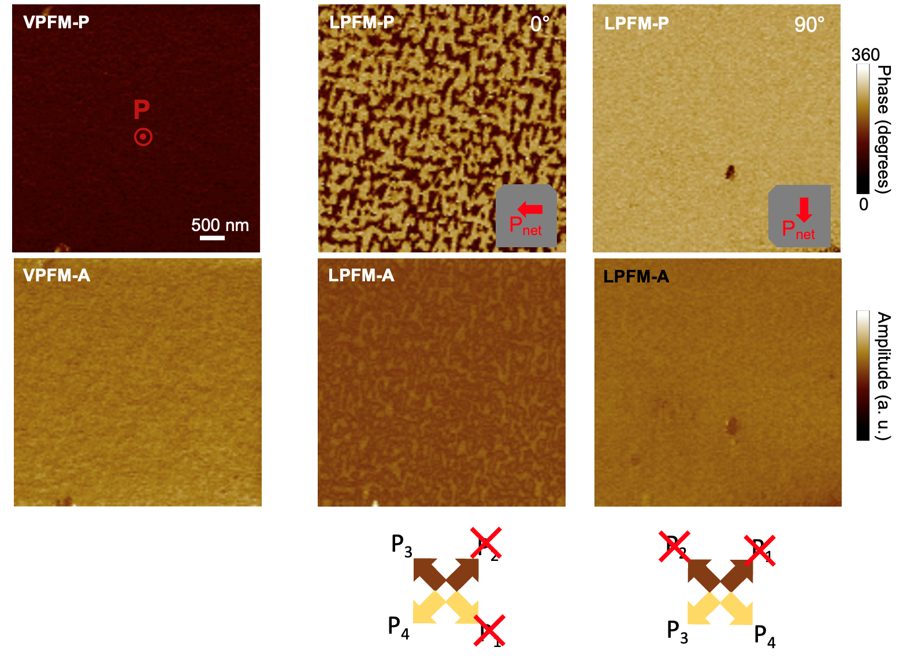}
\caption{PFM characterization of the as-grown state of BiFeO$_3$/DyScO$_3$(110)$_o$. The vertical-PFM phase (VPFM-P) and amplitude (VPFM-A) indicate a uniform polarization pointing upward. The lateral-PFM phase (LPFM-P) and amplitude (LPFM-A) are collected under different angles (90$^\circ$, 0$^\circ$) of the PFM cantilever with respect to the sample to discard the possible in-plane variants of BiFeO$_3$ (P1, P2, P3, P4). The LPFM-P image at 0$^\circ$ shows a disordered stripe-domain pattern, with mainly two alternating ferroelectric variants. For an angle of 90$^\circ$, the LPFM-P becomes uniform and bright, indicating that only the P3 and P4 variants are present. Consequently, vector PFM imaging enables us to know that P3 and P4 are present in the film, with 71$^\circ$ domain walls and the corresponding net in-plane polarization Pnet, as displayed in the sketch (red arrow). In all the PFM images, the cantilever is horizontal and pointing to the left.}\label{fig:laf6}
\end{figure}

The epitaxial thin film of BiFeO$_3$(001)$_{pc}$ was grown by pulsed laser deposition using a KrF excimer laser on a DyScO$_3$(110)$_o$ single crystal substrate without any buffer electrode. The BiFeO$_3$(001)$_{pc}$ film (21 nm) was subsequently grown at the same temperature under 0.36 mbar of oxygen with 1 Hz repetition rate. The sample was slowly cooled down under high oxygen pressure. PFM characterization (Figure~S\ref{fig:laf6}) reveals a homogeneous upward polarization (VPFM phase and amplitude) with mainly two ferroelectric variants self-organized in the form of disordered stripes (LPFM-P at 0$^\circ$).

\begin{figure}[!h]
\includegraphics[width=\columnwidth]{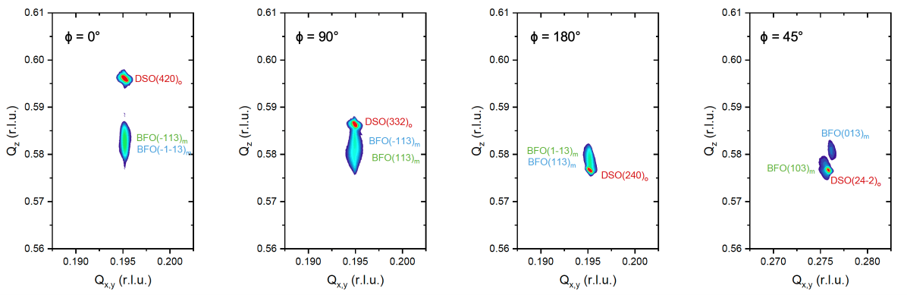}
\caption{X-ray diffraction reciprocal space maps around the (103)$_m$, (013)$_m$ (-103)$_m$ and (113)$_m$ peaks of DyScO$_3$. The BiFeO$_3$ film is fully strained ($\phi$ = 0$^\circ$, 90$^\circ$, 180$^\circ$) by the DyScO$_3$ substrate, as emphasized by the same in-plane Q$_{x,y}$ values between DyScO$_3$ and BiFeO$_3$. Two ferroelastic monoclinic variants of BiFeO$_3$ can be distinguished as emphasized for $\phi$ = 45$^\circ$ and the vertical Q$_z$ splitting for $\phi$= 90$^\circ$. These results are in agreement with a stripe domain sample of alternating ferroelastic monoclinic variants, in which the polarization is rotating by 90$^\circ$ in the film plane.}\label{fig:laf7}
\end{figure}

The corresponding ferroelastic variants associated to these ferroelectric domains were detected using X-ray diffraction reciprocal space maps (Figure~S\ref{fig:laf7}). Reciprocal space maps at $\phi$= 0$^\circ$, 90$^\circ$, 180$^\circ$ show that the BiFeO$_3$ film is fully strained by the DyScO$_3$ substrate. Maps at $\phi$= 90$^\circ$ and 45$^\circ$ show two peaks corresponding to the two mainly monoclinic variants of BiFeO$_3$ arranged with 71$^\circ$ domain walls.

\section{THz transients of all the multidomain BiFeO$_3$ thin films}
\begin{figure}[!ht]
\includegraphics[width=\columnwidth]{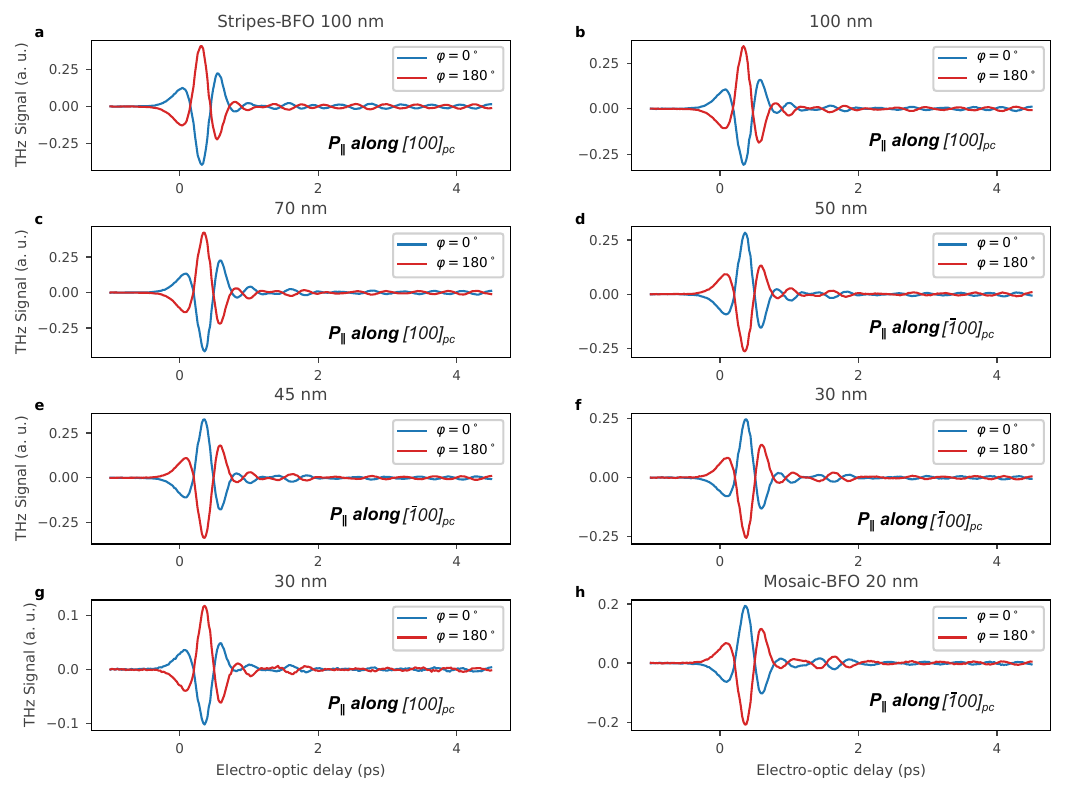}
\caption{\textbf{THz emission from multidomain BiFeO$_3$ thin films excited by an ultrafast optical pulse.} THz waveforms generated at a fluence of $\approx$0.5~mJ/cm$^2$ for each thin film. The pump light pulse is polarized along the $y$-axis ($\alpha=90^\circ$) for either an azimuthal $\varphi$ angle set to $0^\circ$ (blue) or $180^\circ$ (red), while the THz electric field polarization component projected along the laboratory $x$-axis is detected. Note the differing $y$-axis scales.}\label{fig:SM_transients}
\end{figure}
Fig.~S\ref{fig:SM_transients} is showing that the polarity of the THz field is tied to the in-plane ferroelectric macroscopic polarization $\mathbf{P}_\myparallel$ direction.

\section{Maximum signal versus pumping Fluence at 400nm}
The fluence $F$ is define by $F=\dfrac{E}{\pi \omega_0^2}$, with $E$ the pulse energy and $\omega_0$ the beam waist. The fluence data of the Oblique-BFO was fitted with a saturation law, where the saturation fluence was extracted to be $F_0=1.5$mJ/cm$^2$. 
\begin{figure}[!ht]
\includegraphics[width=\columnwidth]{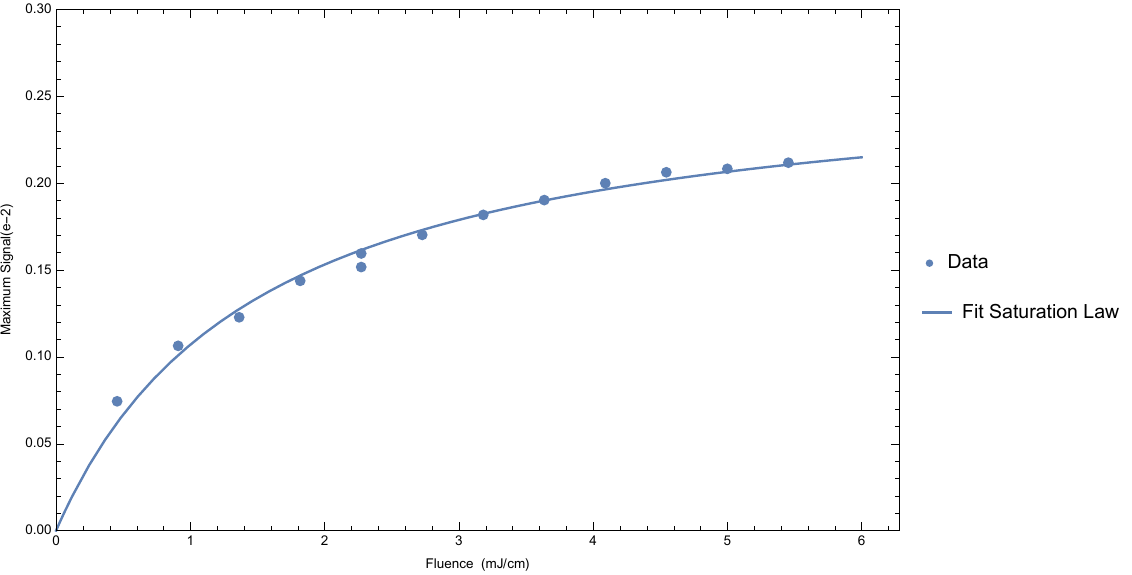}
\caption{Maximum of the signal for the Oblique-BFO (dots). Fitting by a saturation law is represented in the blue line.}
\end{figure}
The other two samples fluence data do not show any saturation behavior and evolve linearly as a function of the pumping fluence. The linear growth rate was found to be $0.4\times10^{-2}$/mJ and $0.1\times10^{-2}$/mJ for the Stripe-BFO and Out-of-Plane-BFO, respectively.

\begin{figure}[!ht]
\includegraphics[width=\columnwidth]{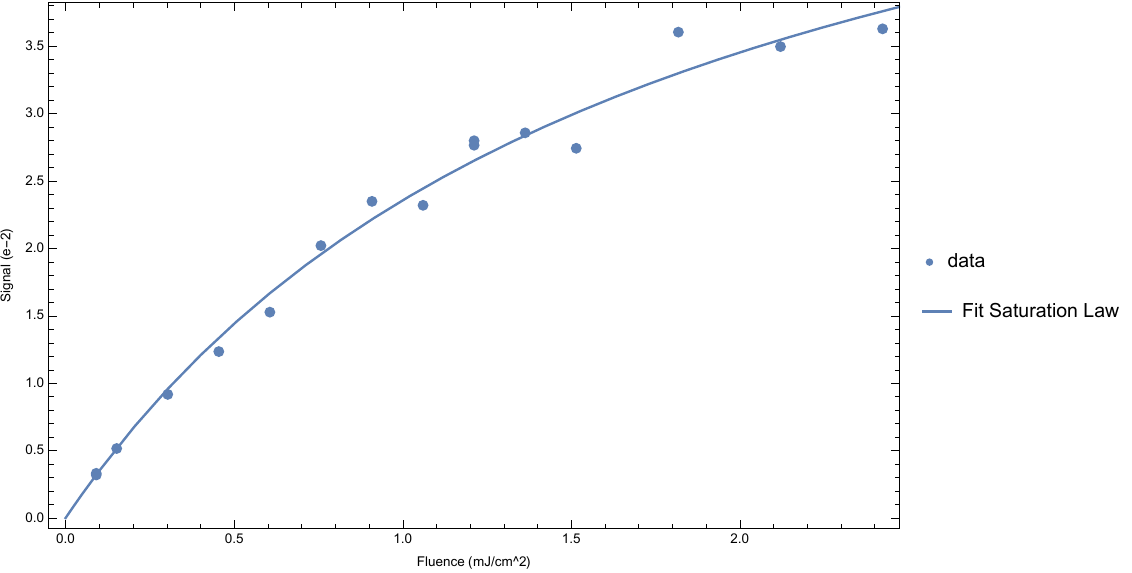}
\caption{Maximum of the signal for the FeCoB/Pt (dots). Fitting by a saturation law is represented in the blue line.}\label{fig:FeCoBPt}
\end{figure}

The fluence data of the FeCoB(5nm)/Pt(3nm) at a pump wavelength 400~nm was fitted with a saturation law, where the saturation fluence was extracted to be $F_0=1.75$mJ/cm$^2$, in line with literature~\cite{Seifert2017,Seifert2022, Vogel2022}. It has to be noted that the data shown in Fig.~S\ref{fig:FeCoBPt} has been taken in air. That's explain the small difference in amplitude with Fig.~3 of the main manuscript. Nevertheless it does not affect the saturation fluence. 

\section{Fourier spectrum of the the different samples investigated in this work.}

\begin{figure}[!ht]
\includegraphics[width=\columnwidth]{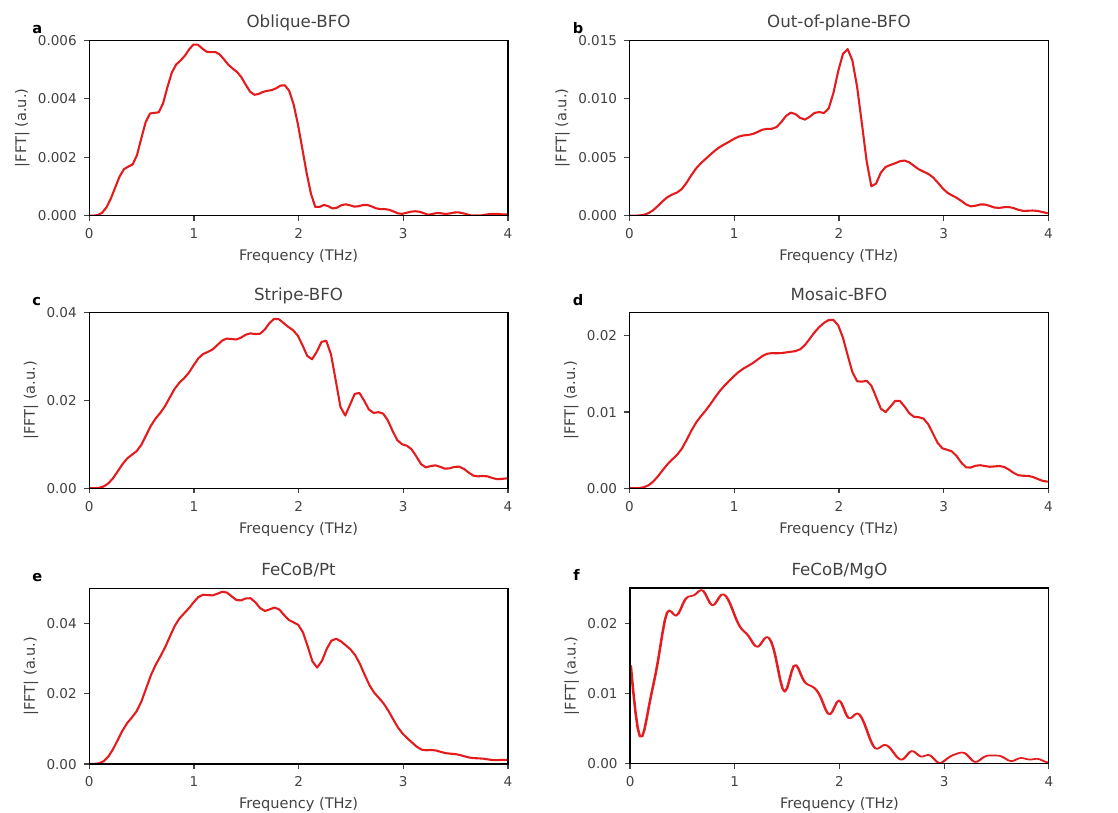}
\caption{FFT of the different samples investigated in this work.}
\end{figure}

\section{Pump wavelength of 800nm: below band gap excitation}

\begin{figure}[!ht]
\includegraphics[width=\columnwidth]{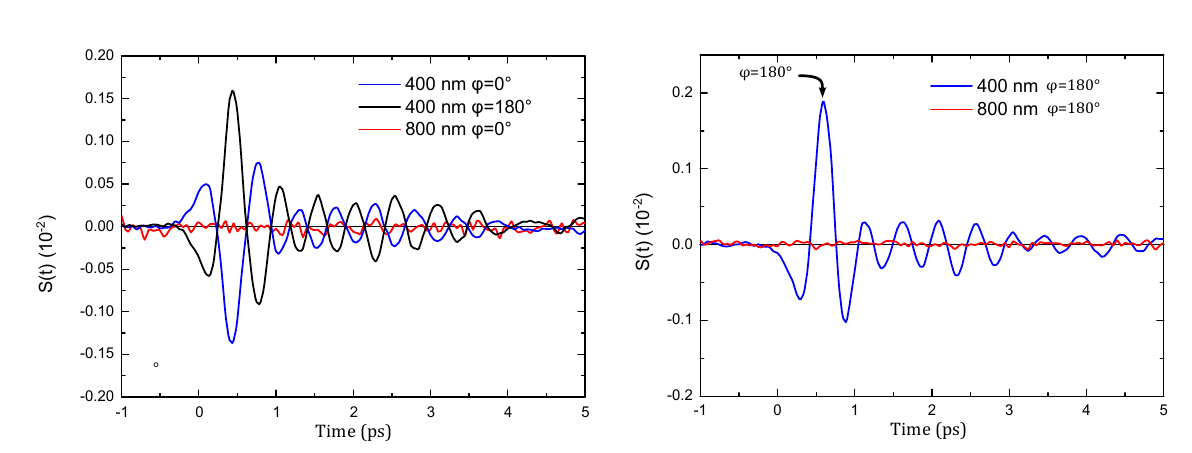}
\caption{Signal recorded at a wavelength of 400~nm (blue and black line) and 800~nm (red line) for a comparable fluence of $\approx2$mJ/cm$^2$ for Oblique-BFO (left) and for the Out-of-Plane-BFO (right).}\label{fig:800nm}
\end{figure}

Fig.~S\ref{fig:800nm} demonstrate that non-resonant second order nonlinearities are negligible in BiFeO$_3$ thin films.

\section{Light-induced transient magnetization in BiFeO$_3$}

Magnetization-related contributions, $\mathbf{J}^\textrm{M} = \mathrm{rot}\left(\mathbf{M}\right)$~\cite{Beaurepaire2004} represents the current generated by a light-induced transient magnetization, which, in the case of BiFeO$_3$, is governed by the spin cycloid~\cite{Fishman2018}. The cycloid is associated with the non-compensated magnetization $\mathbf{M}$ in G-type antiferromagnetic BiFeO$_3$ with spin canting and Dzyaloshinskii-Moriya interaction. Thus, based on the expression of the spin cycloid magnetization given in~\cite{Meisenheimer2024},
the algebra of $\mathrm{rot}\left(\mathbf{M}\right)$ yields the following expression for $J^\textrm{M}$: 
\begin{equation}\label{rotjm}
J^\textrm{M}=k_c m\cos\left(k_c\cdot r_c\right)\mathbf{e_k}\times \mathbf{e_P}+k_c m_\textrm{DM} \sin\left(k_c\cdot r_c\right)\mathbf{e_P},
\end{equation}
where $\mathbf{e_k}$ and $\mathbf{e_P}$ are vectors collinear with the ferroelectric polarization and $k_c$ respectively, $r_c$ is the spatial coordinate along $k_c$ and $m$ and $m_\textrm{DM}$ are the magnetization contribution due to the canting and Dzyaloshinskii-Moriya interaction, respectively. As such, Eq.~\ref{rotjm} indicates that a non-zero current of magnetic origin at the level of the cycloid period (i.e. near field) could exist in the laboratory detection $xy$-plane. However, in our experiment with a pump beam diameter of about 2.5\,mm, the far-field integration over the spatial coordinate $r_c$ results in an average current $\langle J^\textrm{M} \rangle = 0$. This statement holds not only for single-domain thin films (Oblique-BFO and Out-of-plane-BFO) but also for the Stripe-BFO film, since within a single stripe domain the cycloid oscillates more than two times in our case~\cite{Chauleau2020}. 

\section{Optical absorption of the BiFeO$_3$ thin films at 400nm}\label{sec:opticalabs}
Optical reflectivity ($R$) and transmission ($T$) were measured as a function of the light's polarization angle $\alpha$ for a fixed value of $\varphi=0^\circ$ at 400~nm at normal incidence through the BiFeO$_3$ first. The absorption ($A$) was deduced from the data with $A=1-T-R$. The transmission data is not shown as it is very low ($<1\%$). Oblique-BFO shows an optical birefringence as expected~\cite{Gu2023}. No birefringence is noticeable for the other two thin films. All the three thin films show a close light absorption at 400~nm. 

\begin{figure}[!ht]
\includegraphics[width=\columnwidth]{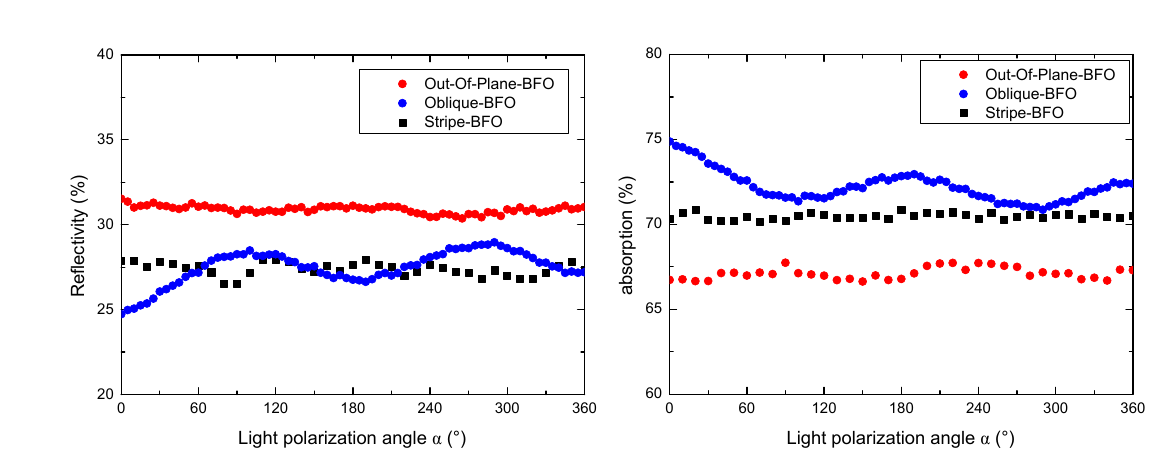}
\caption{400~nm reflectivity (left) and absorption (right) of the samples.}
\end{figure}

\section{Experimental angular dependence}

\subsection{Angular dependence in polar plots}
\begin{figure}[!h]
\includegraphics[width=\columnwidth]{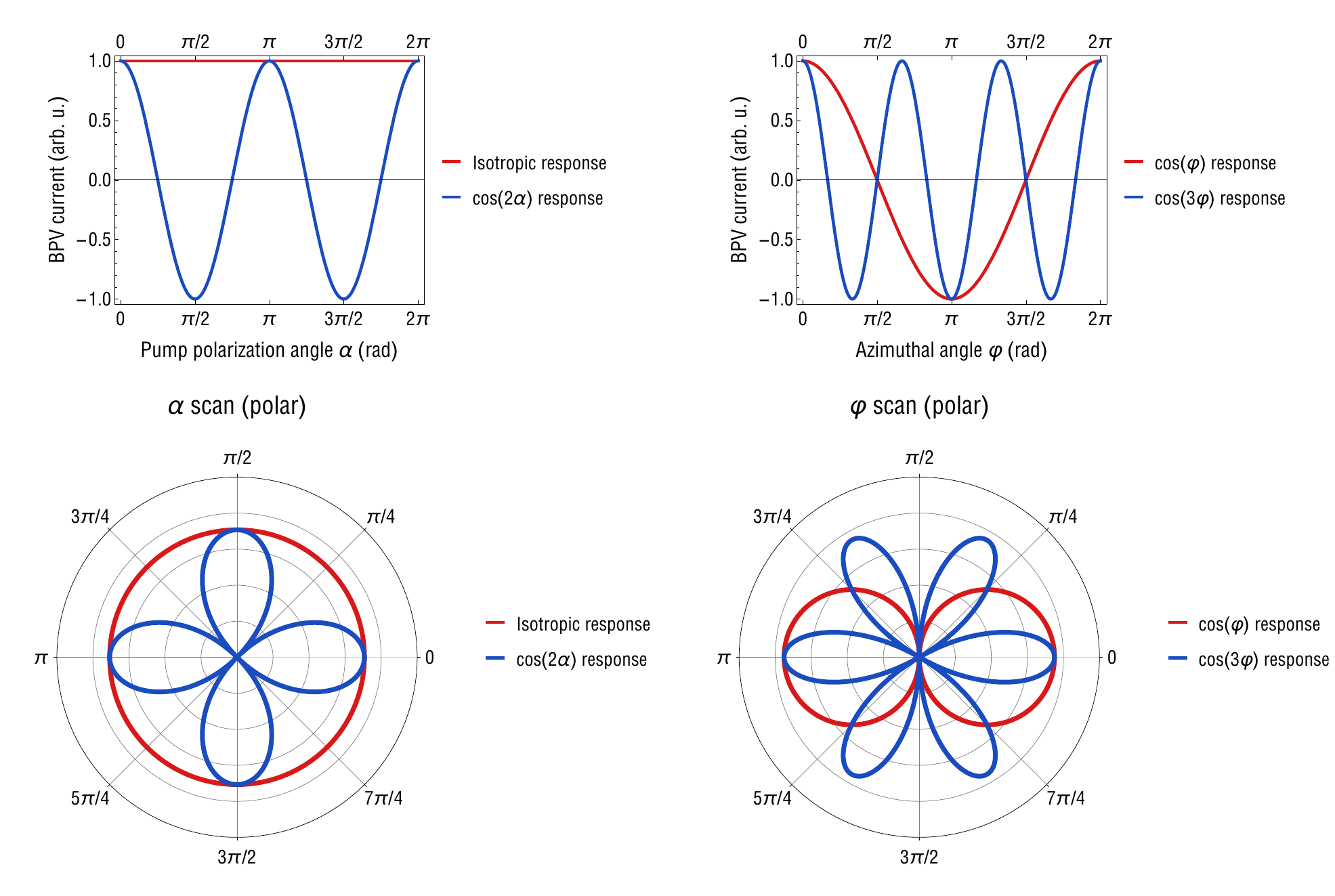}
\caption{Two representations of the isotropic, $\cos (2\alpha)$, $\cos (\varphi)$, and $\cos (3\varphi)$. }\label{fig:sm_representation_out}
\end{figure}

Figure~S\ref{fig:sm_representation_out} displays the polar plots for a set of example angular harmonic (or period) we can encounter in the main manuscript: isotropic, $\cos (2\alpha)$, $\cos (\varphi)$, and $\cos (3\varphi)$.

\subsection{Distinguishing a $\cos2\alpha$ from a $\cos4\alpha$ angular dependence}

The polar plots of the FFT amplitude as a function of the pump polarization angle $\alpha$ exhibit, depending on the sample and frequency, either a two-lobed or a four-lobed (cross-like) shape. Both patterns arise from a single underlying form: a two-fold ($\cos2\alpha$) response superimposed on an isotropic background, whose modulus produces different shapes depending on the relative weight of the two contributions.

\begin{figure}[!h]
\includegraphics[width=\columnwidth]{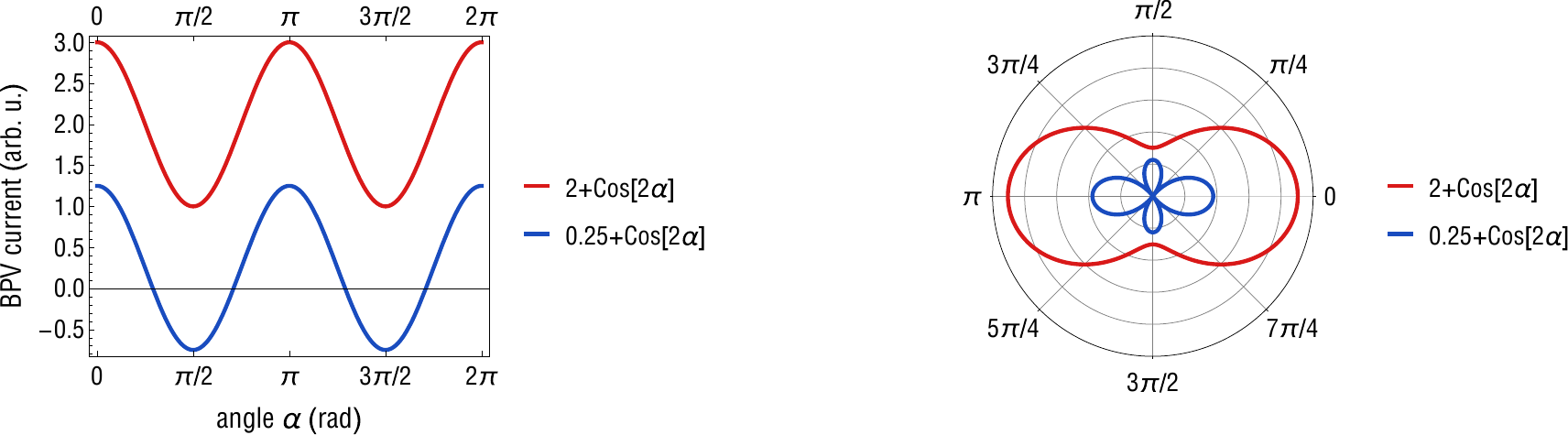}
\caption{Two representations of the isotropic, $\cos (2\alpha)$.}\label{fig:sm_phasepolar_out}
\end{figure}

The measured FFT amplitude is a non-negative quantity. For a signed response of the form $A + B\cos2\alpha$, taking $A, B > 0$ without loss of generality, the polar plot of its modulus $\left|A + B\cos2\alpha\right|$ depends on the ratio $A/B$. When $A/B > 1$, the signed sum never changes sign and the modulus reproduces a two-lobed ``peanut'' shape with a non-zero waist. When $A/B < 1$, the signed sum changes sign at four angles located symmetrically around $\alpha = 90^\circ$ and $270^\circ$; taking the absolute value folds these negative excursions, producing a four-lobed (cross-like) pattern whose petals reach zero, with unequal petal pairs unless $A \ll B$. The transition between the two shapes is governed solely by the relative weight of the anisotropic and isotropic contributions (Fig.~S\ref{fig:sm_phasepolar_out}).



\section{$\mathbf{J}^\textrm{Ani.}$ in BiFeO$_3$: $2\alpha$ and $3\varphi$ harmonics in experimental data}

This section is dedicated to understand more deeply the data and simulation shown in Fig.~4 and~5 of the Main Manuscript.

\subsection{Sample azimuthal rotation $\varphi$-scan in time-domain}
\begin{figure}[!h]
\includegraphics[width=\columnwidth]{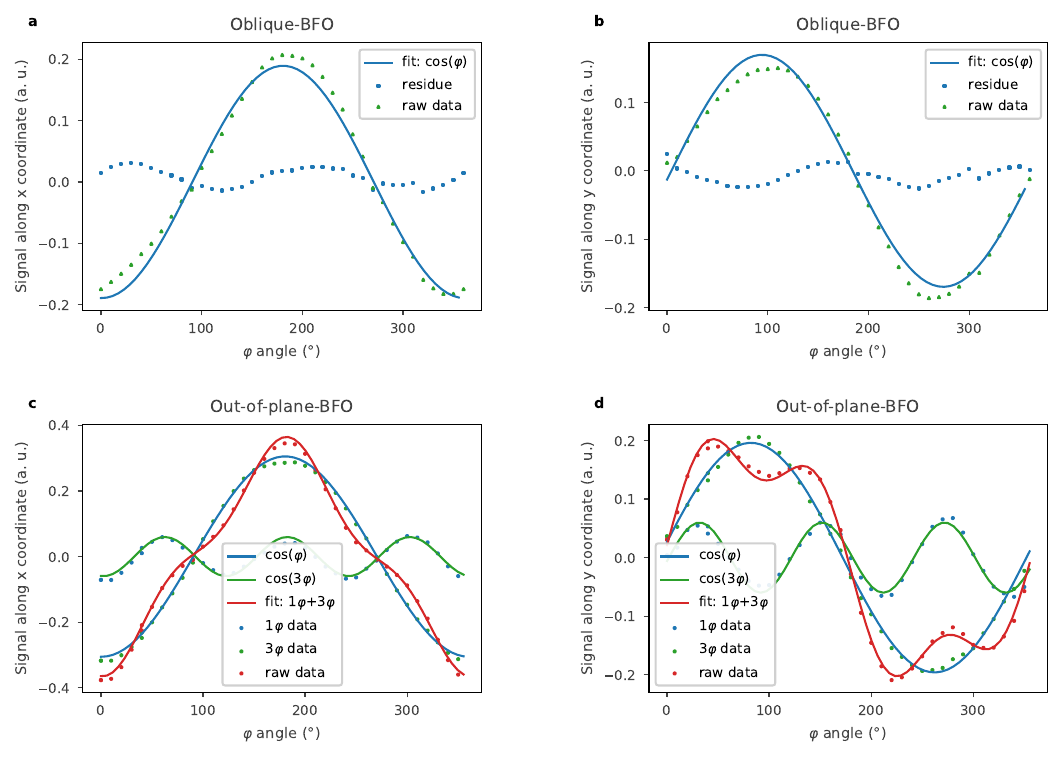}
\caption{Oblique-BFO (top) and Out-of-plane-BFO (bottom) harmonics breakdown. Lines are result of the fit with the harmonics. }\label{fig:SMextrated}
\end{figure}

Figure~S\ref{fig:SMextrated} shows the equivalent of Fig.~4 of the main manuscript. In this case, we break down the different contribution to the raw data to explicitly show the $1\varphi$ and $3\varphi$ harmonics. For Oblique-BFO, the residue shows a small $2\varphi$, which corresponds to the change of photoexcited carriers density due to BiFeO$_3$ birefringence. Birefringence is absent for the other thin films (see section~\ref{sec:opticalabs}).

\subsection{Sample azimuthal rotation $\varphi$-scan in time-domain for different pump polarization angle $\alpha$}
\begin{figure}[!h]
\includegraphics[width=\columnwidth]{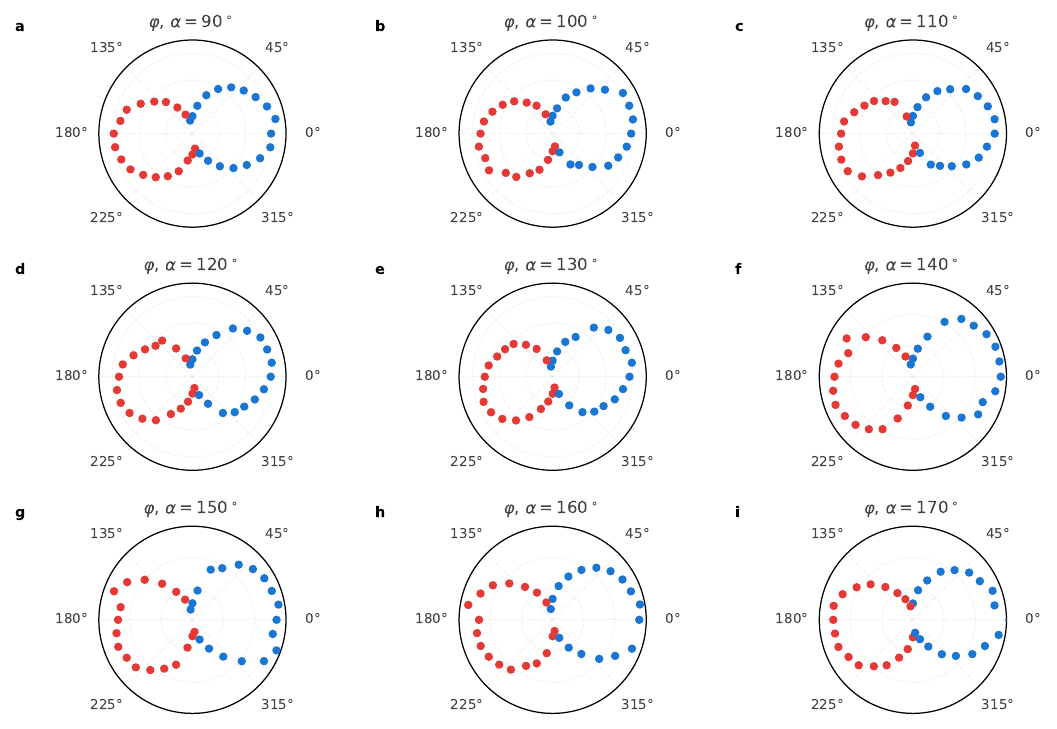}
\caption{\textbf{Angle-resolved THz emission from BiFeO$_3$ thin films: $\varphi$-scan.} Sample azimuthal angle ($\varphi$) dependence of the peak THz field amplitude in the time domain for $x$-detection (dots). Color-encoded by sign (red, positive; blue, negative). Corrected data for the DyScO$_3$ substrate birefringence.}\label{fig:SM2320time}
\end{figure}

Figure~S\ref{fig:SM2320time} shows some polar plots $\varphi$-scan for different different pump polarization angle $\alpha$ for Mosaic-BFO. Other multi-domains samples are equivalent.

\subsection{Sample azimuthal rotation $\varphi$-scan in frequency-domain for different pump polarization angle $\alpha$}
\begin{figure}[!h]
\includegraphics[width=\columnwidth]{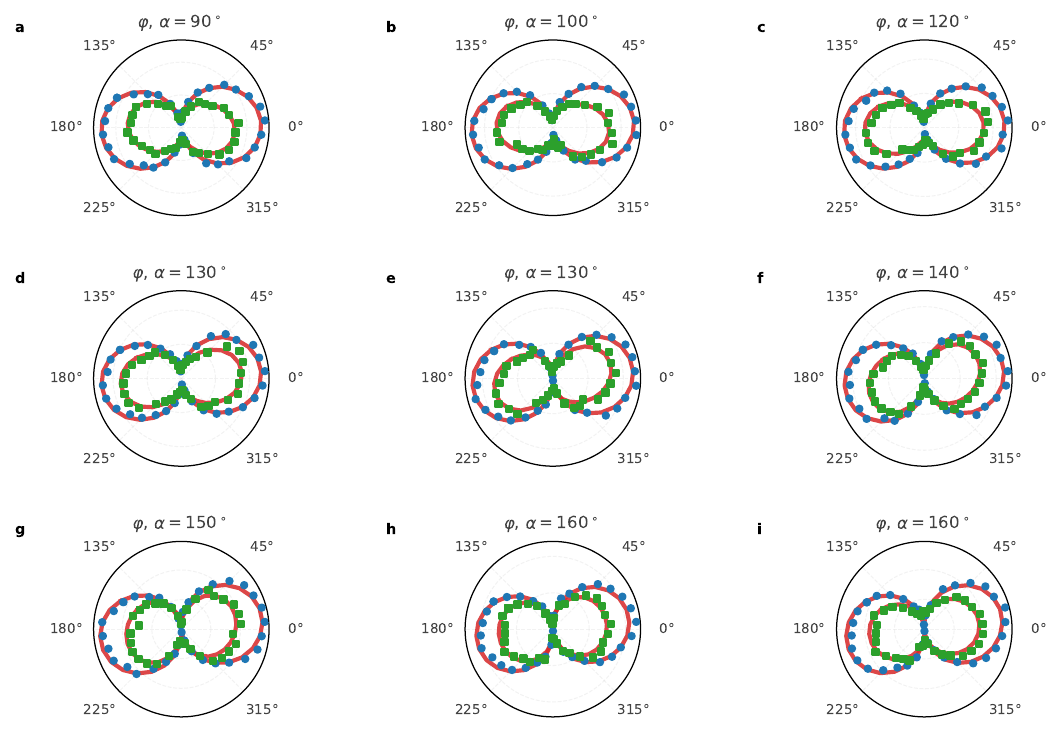}
\caption{Polar plots $\varphi$-scan for different different pump polarization in frequency domain for Mosaic-BFO. Blue and green dots are for a frequency of 1.0 and 2.6 THz, respectively. Lines are the results of our model with $1\varphi$ harmonic. Corrected data for the DyScO$_3$ substrate birefringence.}\label{fig:SM2320freq}
\end{figure}

Figure~S\ref{fig:SM2320freq} shows some polar plots $\varphi$-scan for different different pump polarization angle $\alpha$ for Mosaic-BFO. Other multi-domains samples are equivalent.

\subsection{Light's polarization angle $\alpha$-scan in time-domain}
\begin{figure}[!h]
\includegraphics[width=\columnwidth]{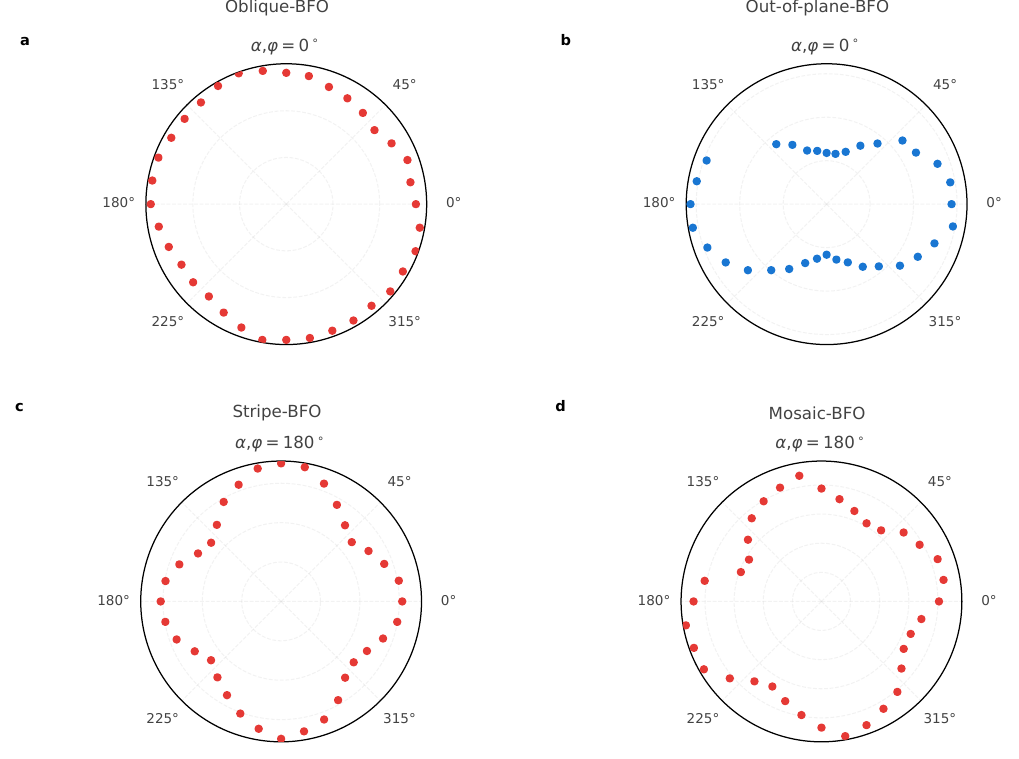}
\caption{ \textbf{Angle-resolved THz emission from BiFeO$_3$ thin films: $\alpha$-scan.} Light's polarization angle ($\alpha$) dependence of the peak THz field amplitude in the time domain $x$-detection (circles), color-encoded by sign (red, positive; blue, negative). Each panel is a polar plot for corresponding to Oblique-BFO, Out-of-plane-BFO, Stripe-BFO, and Mosaic-BFO. Raw data, uncorrected for the DyScO$_3$ substrate birefringence.}\label{fig:SM_alpha_time}
\end{figure}

Figure~S\ref{fig:SM_alpha_time} shows the equivalent of Fig.~4 for $\alpha$-scan.

\section{Bulk photo-voltaic effect in BiFeO$_3$}
The phenomenological expression of the bulk photo-voltaic current $J^\textrm{BPV}$ generation upon excitation of a non-centrosymmetric material by light radiation $E(\omega)$ is given by:
\begin{equation}
    J^\textrm{BPV}_i=\beta^\textrm{BPV}_{i,j,k}E_j(\omega)E_k(\omega),
\end{equation}
where the sum over the repeated indices is omitted.
The rank 3 tensor is defined by:
\[\beta=
\left(\begin{array}{@{}ccc|ccc|ccc@{}}
\beta _{111} & \beta _{112} & \beta _{113} & \beta _{211} & \beta _{212} & \beta _{213} & \beta _{311} & \beta _{312} & \beta _{313}\\
 \beta _{121} & \beta _{122} & \beta _{123} & \beta _{221} & \beta _{222} & \beta _{223} & \beta _{321} & \beta _{322} & \beta _{323} \\
 \beta _{131} & \beta _{132} & \beta _{133} & \beta _{231} & \beta _{23,} & \beta _{233} & \beta _{331} & \beta _{332} & \beta _{333}
\end{array}\right)
\]

The construction of the irreducible form of the tensor is based on the fact that the tensor should remain the same after the use of symmetries operations ($a$) relevant to the point group symmetry. Thus, :
\begin{equation}
    \beta_{ijk}=a_{i,q}a_{j,r}a_{k,s}\beta_{qrs},\label{eq:reduced}
\end{equation}
where the sum over the repeted indices is omitted. Eq.~\ref{eq:reduced} gives conditions on the different elements of $\beta_{ijk}$.

For material with 3m (R3c) point group symmetry, the generating matrix are the 3-fold rotation around the z-axis ($\sigma^\textrm{3z}$) and the mirror plane, which is reflection in a mirror normal to the x-axis ($\sigma^{\over{\textrm{x}}}$):

\begin{equation}
\sigma^\textrm{3z}=\left(
\begin{array}{ccc}
 -\frac{1}{2} & -\frac{\sqrt{3}}{2} & 0 \\
 \frac{\sqrt{3}}{2} & -\frac{1}{2} & 0 \\
 0 & 0 & 1 \\
\end{array}
\right)
\end{equation}
and
\begin{equation}
\sigma^{\over{\textrm{x}}}=\left(
\begin{array}{ccc}
 -1 & 0 & 0 \\
 0 & 1 & 0 \\
 0 & 0 & 1 \\
\end{array}
\right)
\end{equation}

Upon symmetry operation, the BPV tensor in R3c is then expressed by:
\[\beta^\textrm{BPV}=
\left(\begin{array}{@{}ccc|ccc|ccc@{}}
0 & -\beta _{222} & \beta _{113} & -\beta _{222} & 0 & 0 & \beta _{311} & 0 & 0\\
 -\beta _{222} & 0 & 0 &0 & \beta _{222} & \beta _{113} & 0 & \beta _{311} & 0  \\
 \beta _{131} & 0 & 0 & 0 & \beta _{131} & 0 & 0 & 0 & \beta _{333} \\ 
\end{array}\right).
\]

For reference, the calculated $\beta_{ijk}$ values are given in~\cite{Young2012} and, for a wavelength of 400 nm, are $\beta_{311}= 1.6\times10^{-4}V^{-1}$, $\beta_{222}=2.1\times10^{-4}V^{-1}$, $\beta_{131}=\beta_{113}=0.3\times10^{-4}V^{-1}$, $\beta_{333}=3.3\times10^{-4}V^{-1}$. 

In the laboratory frame, the expression of the pump light electric field propagating along the $z$-axis is given by $\mathbf{E}=E_\textrm{Inc.}\exp\left(i(kz-\omega t)\right)\cos\left(\alpha\right) \mathbf{e_x}+E_\textrm{Inc.}\exp\left(i(kz-\omega t)\right)\sin\left(\alpha\right)\mathbf{e_y}$. The hexagonal unit cell axis is usually defined in pseudocubic coordinates by:
\begin{equation}
\left\{
\begin{aligned}
x'_h     &=\dfrac{1}{\sqrt{2}} [1\bar{1}0]_{pc} \\
y'_h    &=\dfrac{1}{\sqrt{6}} [11\bar{2}]_{pc} \\
z'_h   &=\dfrac{1}{\sqrt{3}} [111]_{pc} \\
\end{aligned}
\right.
\end{equation}
where $[001]_h=[111]_{pc}$ is the direction of the ferroelectric polarization.

When the sample coordinate system does not align with the hexagonal unit cell axis, one has to express the BPV tensor in the sample coordinate system by using rotation matrix .

\subsection{Out-of-plane-BFO}

This thin film has its coordinate system expressed by :
\begin{equation}
\left\{
\begin{aligned}
x    &=\dfrac{1}{\sqrt{6}} [1\bar{2}1]_{pc} \\ 
y    &=\dfrac{1}{\sqrt{2}} [10\bar{1}]_{pc} \\
z   &=\dfrac{1}{\sqrt{3}} [111]_{pc} \\
\end{aligned}
\right.
\end{equation}
where $x$, $y$ and $z$ are the laboratory axis coordinate system.

\begin{figure}[!h]
\includegraphics[width=\columnwidth]{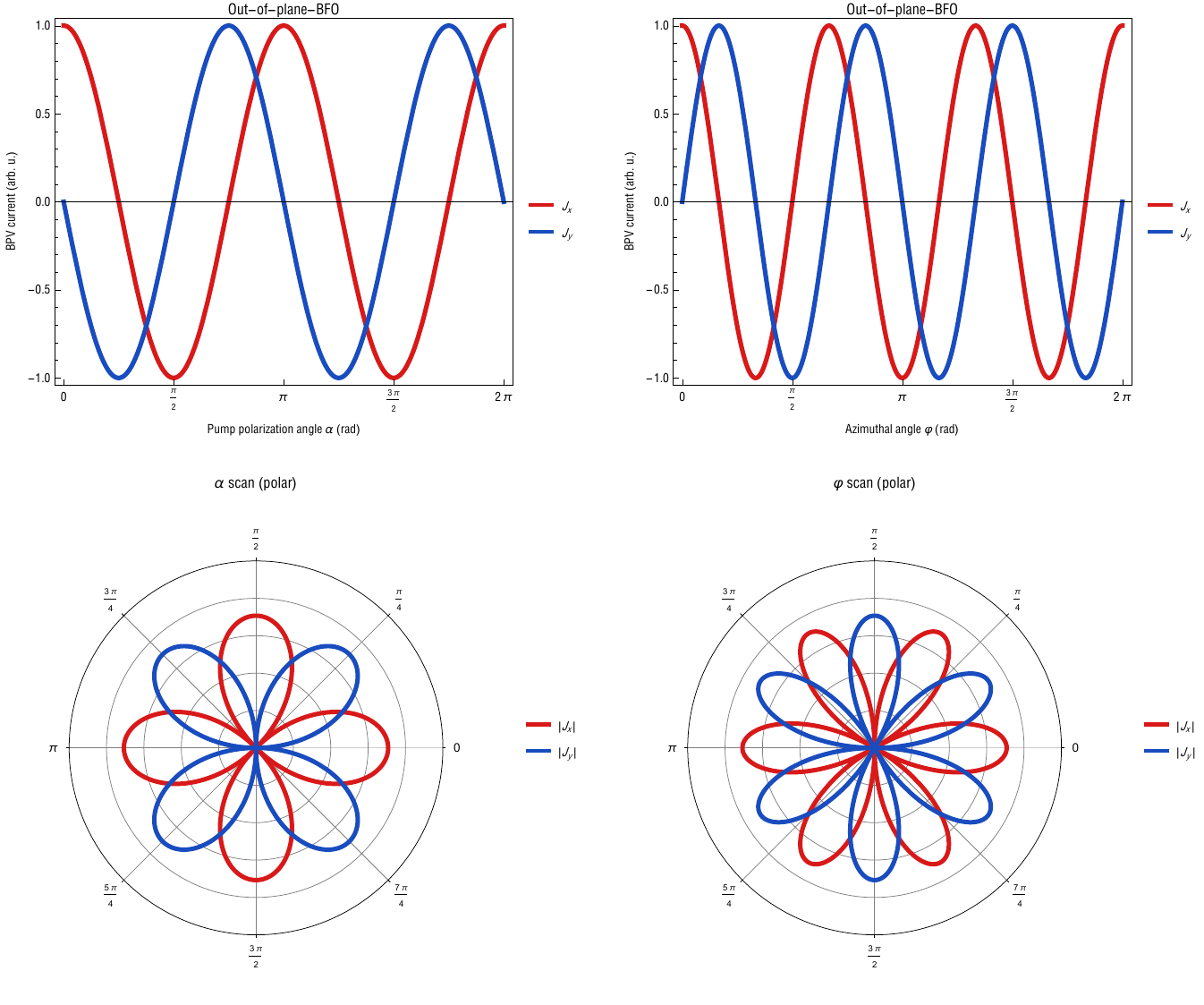}
\caption{Out-of-Plane-BFO BPV current $\textbf{J}^\textrm{BPV}$ in the laboratory frame as a function of the light's polarization angle $\alpha$  (left) and  as a function of the sample azimuthal $\varphi$ (right). $\beta_{222}$ is set to 1$V^{-1}$}\label{fig:sm_bpv_out}
\end{figure}

One can express the BPV current $\boldsymbol{J^\textrm{BPV}}$ as a function of the angle $\alpha$ by in the laboratory frame $(x,y,z)$:
\begin{equation}
\textbf{J}^\textrm{BPV}(\alpha,\varphi)=\left(
\begin{array}{c}
 J^\textrm{BPV}_x \\
 J^\textrm{BPV}_y \\
 J^\textrm{BPV}_z \\
\end{array}
\right)=E_\textrm{Inc.}^2
\left(
\begin{array}{c}
 \beta_{222} \cos (2 \alpha -3 \varphi ) \\
 -\beta_{222} \sin (2 \alpha -3 \varphi ) 
  \\
 \beta_{311} \\
\end{array}
\right).
\end{equation}
The expression of $\boldsymbol{J^\textrm{BPV}}$ is rather simple as the crystallographic orientation of the thin film is epitaxially aligned with the underlying unit cell, such that the film’s lattice axes coincide with those of the crystal structure. We recover the $3m$ symmetry signature with the $3\varphi$ period upon sample rotation. Fig.~S\ref{fig:sm_bpv_out}

\subsection{Oblique-BFO}

This thin film has its coordinate system expressed by:
\begin{equation}
\left\{
\begin{aligned}
x    &=\phantom{\dfrac{1}{\sqrt{2}}}[00\bar{1}] \\ 
y    &=\dfrac{1}{\sqrt{2}} [\bar{1}10] \\
z   &=\dfrac{1}{\sqrt{2}} [110] \\
\end{aligned}
\right.
\end{equation}

\begin{figure}[!h]
\includegraphics[width=\columnwidth]{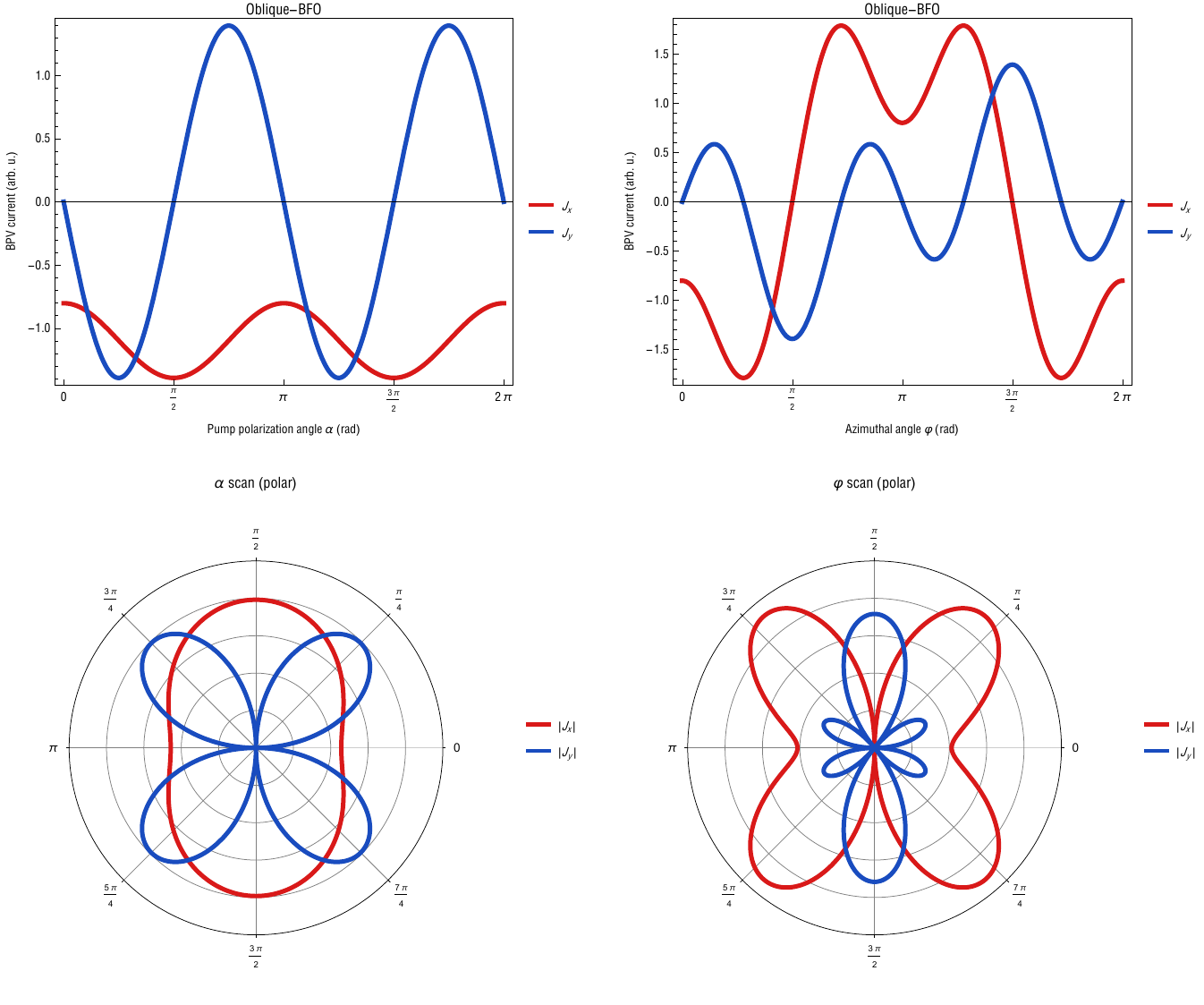}
\caption{Oblique-BFO BPV current $\textbf{J}^\textrm{BPV}$ in the laboratory frame as a function of the light's polarization angle $\alpha$ (left) and  as a function of the sample azimuthal $\varphi$ (right). $\beta_{113}$, $\beta_{222}$, $\beta_{311}$, and $\beta_{333}$ are set to 1$V^{-1}$. }\label{fig:sm_bpv_oblique}
\end{figure}

Following the same reasoning, the BPV current is given by: 
In the case of the Oblique-BFO the BPV current $\boldsymbol{J^\textrm{BPV}}$ as a function of the angles $\alpha$ and $\varphi$ in the laboratory $(x,y)$ plane is:
\begin{equation}
\textbf{J}^\textrm{BPV}(\alpha,\varphi)\propto E_\textrm{Inc.}^2
\left(
\begin{array}{c}
A\cos (2 \alpha -3 \varphi )+B \cos (2 \alpha -\varphi ) + C \cos (\varphi )\\
 -A\sin (2 \alpha -3 \varphi )-B\sin (2 \alpha -\varphi ) +C \sin (\varphi )
 \\
\end{array}
\right).
\end{equation}
$A$, $B$, and $C$ are constants given by combinations of the tensor elements $\beta_{113}$, $\beta_{222}$, $\beta_{311}$, and $\beta_{333}$:
\begin{equation}
\left\{
\begin{aligned}
A&= 2 \beta_{113}+11 \sqrt{2} \beta_{222}+\beta_{311}-\beta_{333}\\
B&=10 \beta_{113}+\sqrt{2} \beta_{222}-\beta_{311}+\beta_{333} \\
C&=2 \left(4 \beta_{113}+\sqrt{2} \beta_{222}+5 \beta_{311}+\beta_{333}\right)
\end{aligned}
\right.
\end{equation}
This equation set is equivalent to the literature~\cite{Yang2017} when $\varphi$ is set to 0. Figure~S\ref{fig:sm_bpv_oblique} displays the BPV current in the laboratory frame.

\subsection{Stripe-BFO and Mosaic-BFO}
This thin film has its coordinate system expressed by:
\begin{equation}
\left\{
\begin{aligned}
x    &=[010]_{pc} \\ 
y    &=[100]_{pc} \\
z   &=[00\bar{1}]_{pc} \\
\end{aligned}
\right.
\end{equation}
Following~\cite{Young2012}, the following transformation matrix for R-type domains is:
\begin{equation}
R=\left(
\begin{array}{ccc}
 \frac{1}{\sqrt{2}} & -\frac{1}{\sqrt{2}} & 0 \\
 \frac{1}{\sqrt{6}} & \frac{1}{\sqrt{6}} & -\sqrt{\frac{2}{3}} \\
 \frac{1}{\sqrt{3}} & \frac{1}{\sqrt{3}} & \frac{1}{\sqrt{3}} \\
\end{array}
\right)
\end{equation}
For L-type domains is:
\begin{equation}
L=\left(
\begin{array}{ccc}
 0 & -\frac{1}{\sqrt{2}} & -\frac{1}{\sqrt{2}} \\
 \sqrt{\frac{2}{3}} & \frac{1}{\sqrt{6}} & -\frac{1}{\sqrt{6}} \\
 \frac{1}{\sqrt{3}} & -\frac{1}{\sqrt{3}} & \frac{1}{\sqrt{3}} \\
\end{array}
\right)
\end{equation}
Following the same reasoning as for the Oblique-BFO and Out-of-Plane-BFO, the BPV current for the Stripe-BFO is given by the average response of the R- and L-type domains:
\begin{equation}\label{transfo}
\begin{aligned}
\boldsymbol{J^\textrm{BPV}}&=(\boldsymbol{J^\textrm{BPV}_R}+\boldsymbol{J^\textrm{BPV}_L})/2
\end{aligned}
\end{equation}

The BPV current is given by: 
\begin{align}
&\textbf{J}^\textrm{BPV}(\alpha,\varphi)\propto E_\textrm{Inc.}^2\times\\
&\left(
\begin{array}{c}
 D \cos (2 \alpha -3 \varphi ) +E\cos (2 \alpha -\varphi )+\sqrt{2} \beta_{222} (5 \sin (2 \alpha -3 \varphi )-\sin (2 \alpha -\varphi )+2 \sin (\varphi ))+F \cos (\varphi ) \\
 -D \sin (2 \alpha -3 \varphi )+E\sin (2 \alpha -\varphi )+\sqrt{2} \beta_{222} (5 \cos (2 \alpha -3 \varphi )+\cos (2 \alpha -\varphi )-2 \cos (\varphi ))+F \sin (\varphi )
\end{array}
\right).
\end{align}
$D$, $E$, and $F$ are constants given by combinations of the tensor elements $\beta_{113}$, $\beta_{222}$, $\beta_{311}$, and $\beta_{333}$:

\begin{equation}
\left\{
\begin{aligned}
D&= 2 \left(2 \beta_{113}+\beta_{311}-\beta_{333}\right)\\
E&=2 \left(4 \beta_{113}-\beta_{311}+\beta_{333}\right) \\
F&=4 \left(\beta_{113}+2 \beta_{311}+\beta_{333}\right)
\end{aligned}
\right.
\end{equation}

\begin{figure}[!h]
\includegraphics[width=\columnwidth]{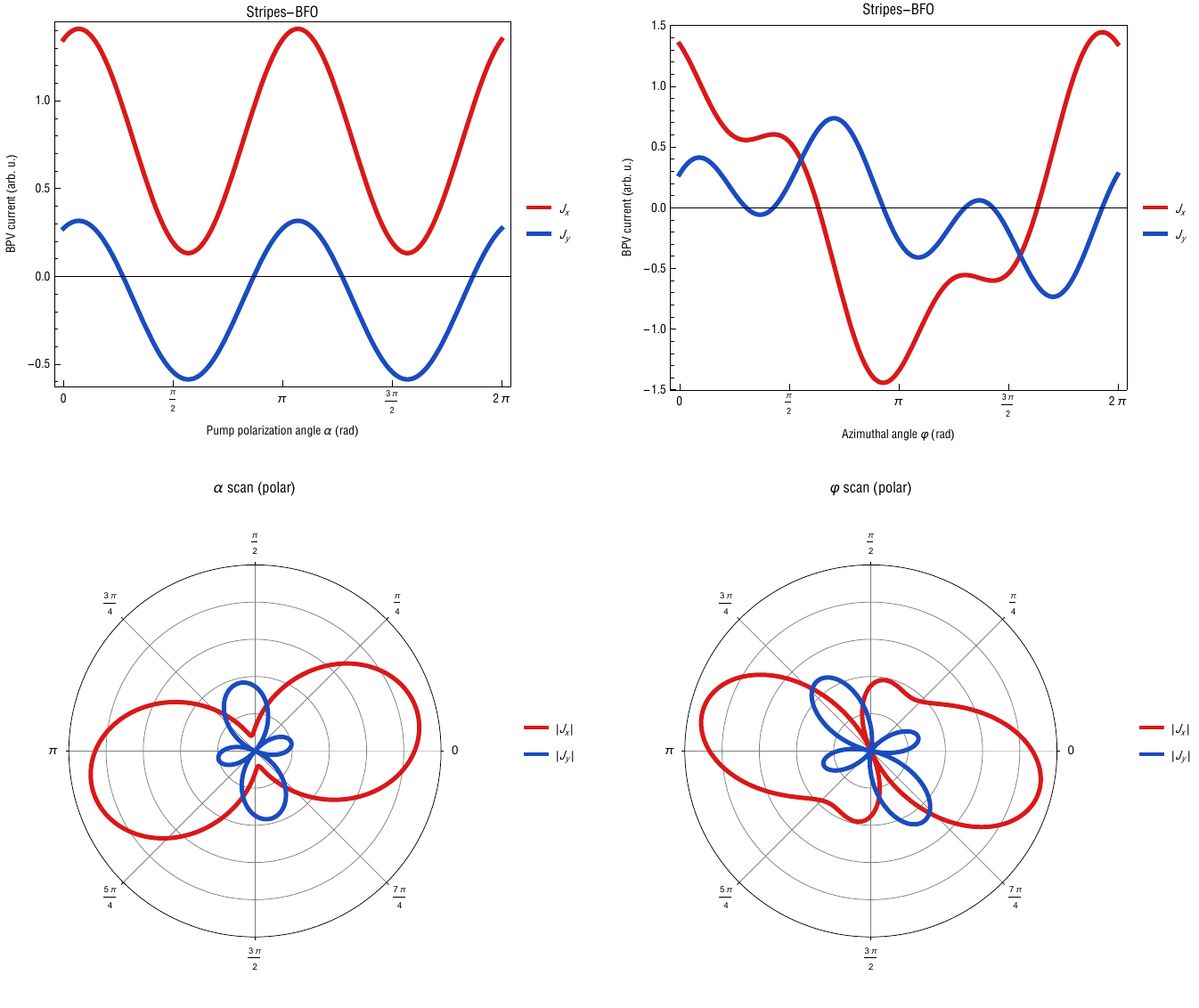}
\caption{Stripe-BFO BPV current $\textbf{J}^\textrm{BPV}$ in the laboratory frame as a function of the light's polarization angle $\alpha$ (left) and  as a function of the sample azimuthal $\varphi$ (right).}
\end{figure}




\section{Carriers acceleration in BiFeO$_3$}
\subsection{Depolarizing electric field collinear to the ferroelectric polarization}

Photo-carriers acceleration by an internal electric field produce a current $\textbf{J}^\textrm{PC}$, which can be expressed in tensorial form by:
\begin{equation}
    \textbf{J}^\textrm{PC}_i=\sigma_{ijkl}E_j(\omega)E_k^\star(\omega)E_l^\mathrm{internal},\label{eq:photo}
\end{equation}
where the sum over the repeated indices is omitted. $E^{\mathrm{internal}}$ is the internal depolarizing electric field. The 4-rank tensor can be further reduced by knowing the fact that the $E^\mathrm{internal}$ is along the [001] direction in the rhombohedral unit cell BiFeO$_3$.
\begin{equation}
    \sigma^\mathrm{Eff.}_{ikl}=\sum_{j=1}^3\sigma_{ijkl}E_j^\mathrm{internal}
\end{equation}
After repeating the symmetry operations stated above, and, as we have only one beam interaction, the indices $j$ and $k$ can be freely permuted in Eq.~\ref{eq:photo}, hence $j,k=k,j$ and $\sigma _{1133}=\sigma _{1313}$. Thus,
\[\sigma^\mathrm{Eff.}=
\left(\begin{array}{@{}ccc|ccc|ccc@{}}
0 & -\sigma _{2223} & \sigma _{1133} & -\sigma _{2223} & 0 & 0 & \sigma _{3113} & 0 & 0\\
 -\sigma _{2223} & 0 & 0 &0 & \sigma _{2223} & \sigma _{1113} & 0 & \sigma _{3113} & 0  \\
 \sigma _{1113} & 0 & 0 & 0 &\sigma _{3113} & 0 & 0 & 0 & \sigma _{3333} \\ 
\end{array}\right),
\]
The general form of the photoconductive current is then given by:
\begin{equation}
    J^\textrm{PC}_i=\left |E_0^\mathrm{internal}\right|\sigma^\mathrm{Eff.}_{ijk}E_j(\omega)E^\star_k(\omega) \label{eq:photo3},
\end{equation}

It can be noted that the tensors $\sigma^{\mathrm{Eff.}}_{ijk}$ and $\beta^{\mathrm{BPV}}_{ijk}$ share the same symmetry elements. This implies that, from a symmetry standpoint, it is not possible to distinguish between $\mathbf{J}^{\mathrm{PC}}$ and $\mathbf{J}^{\mathrm{BPV}}$ when the depolarizing field points along the [111]$_{pc}$ ($[001]_h$) direction, corresponding to the high-symmetry axis of the $3m$ point group. Consequently, given the non-zero tensor elements of $\sigma^{\mathrm{Eff.}}$, the angular dependence of $\mathbf{J}^{\mathrm{PC}}(\alpha,\varphi)$ is \emph{a priori} identical to that of $\mathbf{J}^{\mathrm{BPV}}(\alpha,\varphi)$.

\subsection{Depolarizing electric field perpendicular to the thin film surface}

The electrostatic boundary conditions can in principle alter the direction of the depolarizing field. In monodomain thin films, it may be oriented along the surface normal~\cite{Ghosez2006} rather than along the polarization. For Oblique-BFO, the depolarizing field would then point along $[110]_{\mathrm{pc}}$, no longer collinear with $\mathbf{P}$. In this configuration, the effective photoconductive response $\sigma^{\mathrm{Eff.}}$ acquires additional non-zero elements. The resulting current $\mathbf{J}^\textrm{PC}$ is given by: 
\begin{align}
&\textbf{J}^\textrm{PC}(\alpha,\varphi)\propto E_\textrm{Inc.}^2\times\nonumber\\
&\left(
\begin{array}{c}
 G \cos (2 \alpha -3 \varphi ) +H\cos (2 \alpha -\varphi )+J\cos (\varphi ) + K (-\cos (2\alpha-\varphi)+\cos (\varphi)) \\
 -G \sin (2 \alpha -3 \varphi )+H\sin (2 \alpha -\varphi )+J\sin (\varphi ) - K (\sin (2\alpha-\varphi)-\sin (\varphi))
\end{array}
\right).
\end{align}
$G$, $H$, $J$, ad $K$ are constants given by combinations of the tensor elements $\sigma_{1111}$, $\sigma_{1123}$, $\sigma_{1132}$, $\sigma_{1133}$, $\sigma_{1331}$, $\sigma_{3112}$, $\sigma_{3113}$, $\sigma_{3131}$, and $\sigma_{3333}$. In this case, the harmonics present are: $3\varphi$, $1\varphi$, and $2\alpha$.

\section{Harmonic decomposition}
\subsection{FeCoB(5nm)/Pt(3nm)}
It shows negligible harmonics contribution ($1\alpha$, $2\alpha$ \dots) compare to the isotropic contribution, as expected for this spintronic emitter.
\begin{figure}[!h]
\includegraphics[width=\columnwidth]{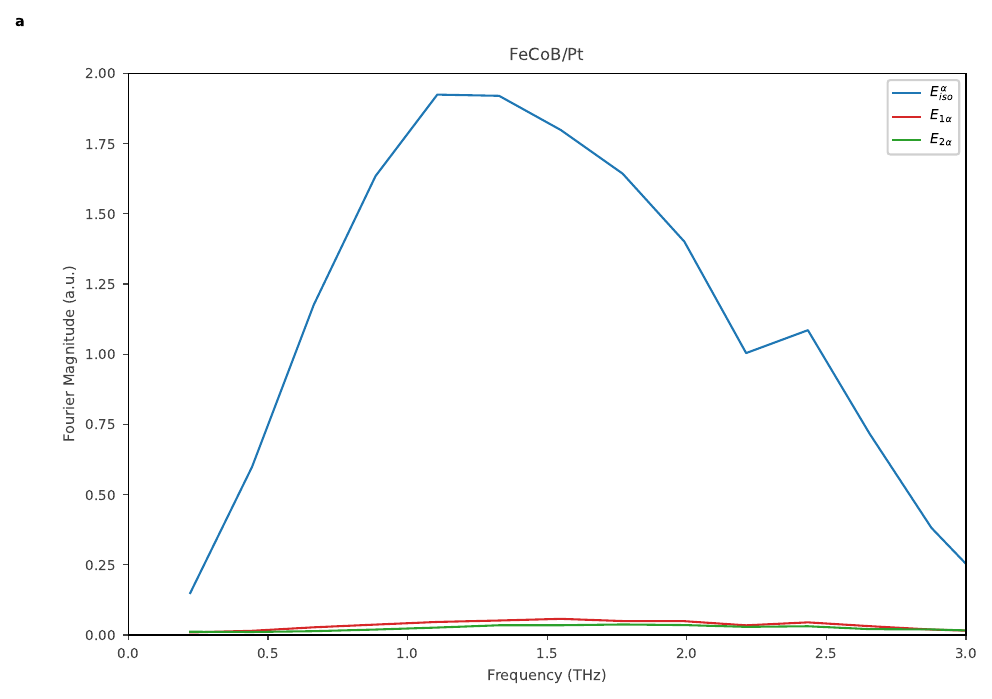}
\caption{Harmonic decomposition for the FeCoB(5nm)/Pt(3nm) emitter in term of light's polarization angle $\alpha$.}
\end{figure}

\subsection{BiFeO$_3$ thin films}
\begin{figure}[!h]
\includegraphics[width=\columnwidth]{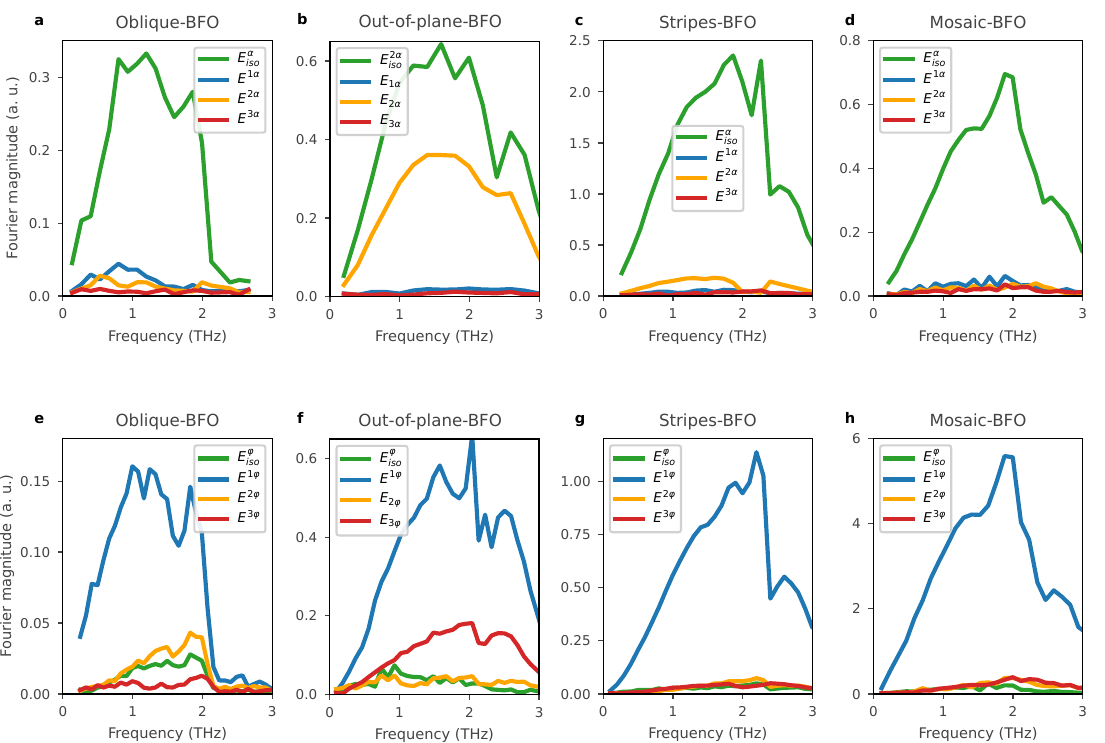}
\caption{All orders for the Harmonic decomposition for the BiFeO$_3$ thin films in term of light's polarization angle $\alpha$ (top) and sample azimuthal angle $\varphi$ (bottom). Corrected data for the DyScO$_3$ substrate birefringence. }\label{fig:sm_harmonics}
\end{figure}

Figure~S\ref{fig:sm_harmonics} shows the complete harmonic decomposition for the light's polarization angle $\alpha$ and the sample azymuthal angle $\varphi$. It shows negligible contribution of the harmonics which are not symmetry-allowed (see Table~1 in the main manuscript). 

\begin{figure}[!h]
\includegraphics[width=\columnwidth]{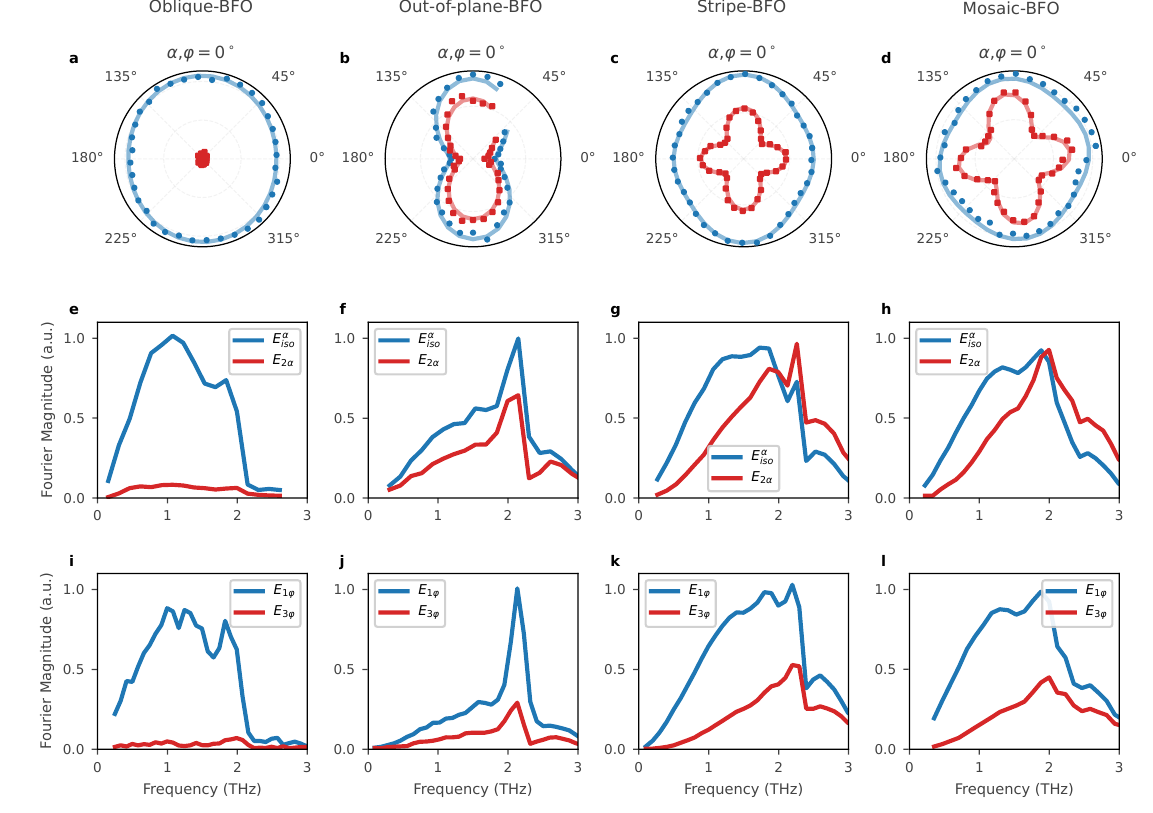}
\caption{\textbf{Effect of the substrate pulse-splitting on the harmonics decomposition.} Columns correspond to Oblique-BFO, Out-of-plane-BFO, Stripe-BFO and Mosaic-BFO. (a--d)~Angular dependence of the emitted field amplitude at 1.0 (blue dots) and \SI{2.6}{THz} (red dots) on the pump polarization angle $\alpha$; lines are the harmonic decomposition into isotropic and $\cos2\alpha$ components. (e--h)~Magnitude of the symmetry-allowed $\alpha$-harmonics $\tilde{E}^\alpha_\textrm{iso}$ (blue) and $\tilde{E}_{2\alpha}$ (red). (i--l)~Magnitude of the symmetry-allowed azimuthal harmonics $\tilde{E}_{1\varphi}$ (blue) and $\tilde{E}_{3\varphi}$ (red). }\label{fig:figure5_SM_Uncorrected}
\end{figure}

Figure~S\ref{fig:figure5_SM_Uncorrected} shows the equivalent of the Figure~5 of the main manuscript but when the substrate pulse-splitting is not corrected for.
\subsection{Reconstructed THz transients}
\begin{figure}[!h]
\includegraphics[width=\columnwidth]{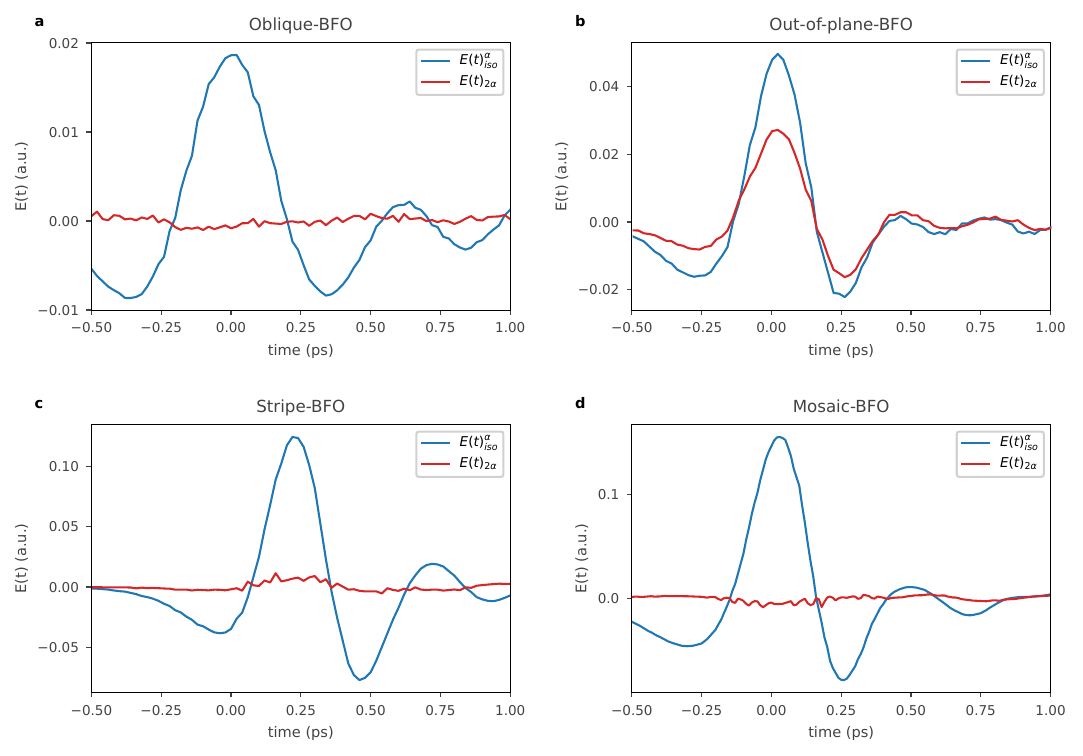}
\caption{Reconstructed THz signal in time-domain for $\tilde{E}(t)^\alpha_\textrm{iso}$ (blue) and $\tilde{E}(t)_{2\alpha}$ (red) for Oblique-BFO (a), Out-of-plane-BFO (b), Stripe-BFO (c), and Mosaic-BFO (d). Corrected data for the DyScO$_3$ substrate birefringence. }\label{fig:sm_recon}
\end{figure}
\section{Substrate birefringence: pump-pulse splitting and its correction}
\label{si:walkoff}
 
\begin{figure}[t]
  \centering
   \includegraphics[width=\linewidth]{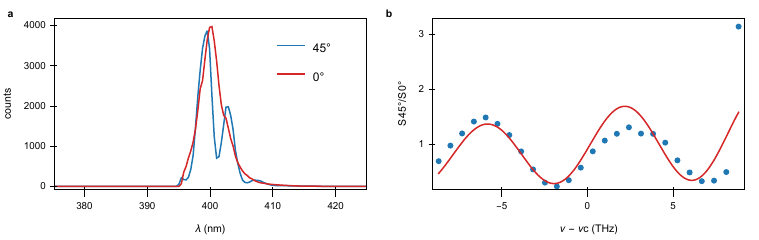}
  \caption{\textbf{Substrate birefringence: Spectral interferometry on the \SI{400}{nm} pump pulse for DyScO$_3$(110).}
  (a) \SI{400}{nm} pump pulse spectra with the substrate fast or slow axis at 0$^\circ$ and 45$^\circ$ respect from the pump light polarization.
  (b) Ratio of the two spectra exhibiting the fringes. The fit is shown as well in red.}
  \label{fig:si_walkoff_pump}
\end{figure}

\begin{figure}[t]
  \centering
   \includegraphics[width=\linewidth]{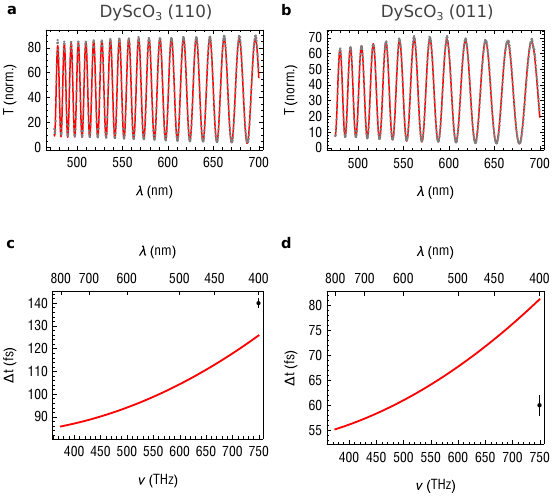}
  \caption{\textbf{Substrate birefringence: White-light channeled spectra for the two DyScO$_3$ substrates.}
  White-light channeled spectra of the bare substrates between parallel polarizers
  (axes at $45^\circ$) and global chirped-cosine fits (top).
  Extracted dispersion
  laws $\Delta t(\nu)$ with the THz-derived values at \SI{400}{nm} (bottom).}
  \label{fig:si_walkoff}
\end{figure}
The orthorhombic DyScO$_3$ substrates are birefringent at the pump wavelength. A pump
pulse polarized at an angle to the substrate principal axes splits into two
orthogonally polarized components arriving at the film with the group delay
$\Delta t = d\,\Delta n_g/c$ ($d$: thickness, $\Delta n_g$: in-plane group-index
difference). For $d = \SI{0.5}{mm}$ and our $\approx\SI{50}{fs}$ pump pulses
(\SI{35}{fs} at \SI{800}{nm}, frequency-doubled in \SI{0.2}{mm}-thick BBO), the film is excited by a
\emph{double pulse} rather than by a single pulse of rotated polarization.
 
\subsection{Group delay}
 
$\Delta t$ was determined by two independent methods (Fig.~S\ref{fig:si_walkoff}a,b).
In situ, from the emission data: the pump-angle dependence of the THz arrival time
follows the double-pulse law $t_c(\alpha) = t_0 + \tfrac{\Delta t}{2}[1-\cos2(\alpha-\psi)]$,
$\psi$ being the substrate fast axis. Independently, from the white-light channeled
spectrum of the bare substrate between parallel polarizers, whose fringe period in
optical frequency measures the group delay directly ($d\delta/d\nu = 2\pi\Delta t$);
its dispersion over \SIrange{450}{750}{nm} is extrapolated to \SI{400}{nm}, a value
confirmed by spectral interference fringes on the pump pulse itself. For
DyScO$_3$(110)$_\mathrm{o}$ the two methods give $\Delta t = \SI{132(2)}{fs}$ (emission
timing) and \SI{125}{fs} (optical), $\Delta n_g \approx 0.075$; for
DyScO$_3$(011)$_\mathrm{o}$ the optical value $\Delta t = \SI{80}{fs}$
($\Delta n_g \approx 0.048$) is adopted, the emission-timing estimate being affected by
the partial overlap of the two components on this weakly birefringent cut. No delay is
found on cubic SrTiO$_3$. The field-envelope overlap of the two components,
$V = 2^{-(\Delta t/\tau)^2}$ (intensity FWHM $\tau$), is $\approx 0.01$ on the
(110)$_\mathrm{o}$ cut, fully separated, and $\approx 0.17$ on the
(011)$_\mathrm{o}$ cut.

\begin{figure}[t]
  \centering
   \includegraphics[width=\linewidth]{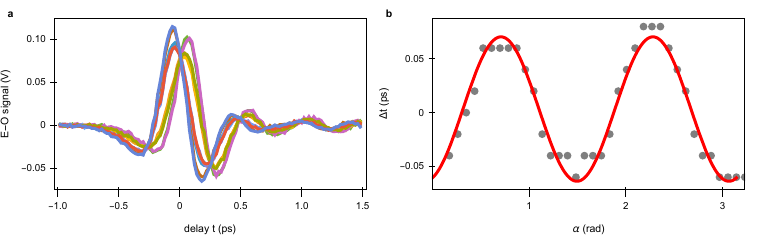}
  \caption{\textbf{Different time of arrival of the THz transients due to the substrate in Stripe-BFO/DyScO$_3$(110).}
  Selected THz transients for Stripe-BFO at different $\alpha$ angles (a).
  Extracted timing position of the peak signal for the correction (b).}
  \label{fig:si_walkoff}
\end{figure}
The orthorhombic DyScO$_3$ substrates are birefringent at the pump wavelength. A pump
pulse polarized at an angle to the substrate principal axes splits into two
orthogonally polarized components arriving at the film with the group delay
$\Delta t = d\,\Delta n_g/c$ ($d$: thickness, $\Delta n_g$: in-plane group-index
difference). For $d = \SI{0.5}{mm}$ and our $\approx\SI{50}{fs}$ pump pulses
(\SI{35}{fs} at \SI{800}{nm}, frequency-doubled in BBO), the film is excited by a
\emph{double pulse} rather than by a single pulse of rotated polarization.
\bibliography{export2}